\documentclass{jfm}

\usepackage{capt-of}

\usepackage{graphicx}
\usepackage{newtxtext}
\usepackage{newtxmath}
\usepackage{natbib}
\usepackage{hyperref}
\usepackage{nth}
\usepackage{caption} 
\hypersetup{
    colorlinks = true,
    urlcolor   = blue,
    citecolor  = black,
}

\newcommand{\RomanNumeralCaps}[1]
\linenumbers

\usepackage[x11names]{xcolor}       
\usepackage{soul}          

\newif\ifrevmode
\revmodetrue          

\ifrevmode
  
\else
  
\fi

\title{Solute dispersion boosts the phoretic removal of colloids from dead-end pores}

\author{Yiran Li\aff{1},
  Mobin Alipour\aff{1}
 \and Amir A. Pahlavan\aff{1}\corresp{\email{amir.pahlavan@yale.edu}}}

\affiliation{\aff{1}Department of Mechanical Engineering and Materials Science,\\
 Yale University, New Haven, Connecticut 06511, USA.}

\begin{document}
\maketitle

\begin{abstract}
Predicting and controlling the transport of colloids in porous media is essential for a broad range of applications, from drug delivery to contaminant remediation. Chemical gradients are ubiquitous in these environments, arising from reactions, precipitation/dissolution, or salinity contrasts, and can drive particle motion via diffusiophoresis. Yet our current understanding mostly comes from idealized settings with sharply imposed solute gradients, whereas in porous media, flow disorder enhances solute dispersion, and leads to diffuse solute fronts. This raises a central question: does front dispersion suppress diffusiophoretic migration of colloids in dead-end pores, rendering the effect negligible at larger scales? We address this question using an idealized one-dimensional dead-end geometry. We derive an analytical model for the spatiotemporal evolution of colloids subjected to slowly varying solute fronts and validate it with numerical simulations and microfluidic experiments. Counterintuitively, we find that diffuseness of solute front enhances removal from dead-end pores: although smoothing reduces instantaneous gradient magnitude, it extends the temporal extent of phoretic forcing, yielding a larger cumulative drift and higher clearance efficiency than sharp fronts. Our results highlight that solute dispersion does not weaken the phoretic migration of colloids from dead-end pores, pointing to the potential relevance of diffusiophoresis at larger scales, with implications for filtration, remediation, and targeted delivery in porous media.

\end{abstract}

\begin{keywords}
Authors should not enter keywords on the manuscript, as these must be chosen by the author during the online submission process and will then be added during the typesetting process (see \href{https://www.cambridge.org/core/journals/journal-of-fluid-mechanics/information/list-of-keywords}{Keyword PDF} for the full list).  Other classifications will be added at the same time.
\end{keywords}

{\bf MSC Codes }  {\it(Optional)} Please enter your MSC Codes here

\section{\label{sec:intro}Introduction}

Predicting and controlling colloid transport in porous media is essential across diverse fields, from targeted drug delivery \citep{manzari_targeted_2021,mitchell_engineering_2021}, to managing the spread of microplastics and contaminants in subsurface environments and coastal aquifers \citep{alimi_microplastics_2018,brewer_mobility_2021,pahlavan_soil_2024,spielman-sun_critical_2024}. Chemical gradients, prevalent in natural and engineered porous systems \citep{dentz_mixing_2011,dentz_mixing_2023,rolle_mixing_2019,berkowitz_measurements_2016}, can induce colloidal motion through diffusiophoresis \citep{anderson_colloid_1989,velegol_origins_2016,marbach_osmosis_2019,shim_diffusiophoresis_2022,ault_physicochemical_2024}. For example, ionic gradients resulting from agricultural practices, such as pesticide application and irrigation, can mobilize microplastics or natural colloids and clay particles \citep{muller_critical_2007}. These mobilized colloids can facilitate the transport of contaminants, impacting groundwater quality, ecosystem health, and subsurface biodiversity \citep{mccarthy_subsurface_1989,de_jonge_colloids_2004,sen_review_2006,weber_contaminant_2009,sheng_plastic_2024}. A critical open question is whether solute gradients significantly influence the macroscopic transport of colloids in these complex environments.

Our understanding of the fundamental aspects of solute-mediated migration of colloids in idealized scenarios characterized by sharp solute gradients and no background flows, e.g. in dead-end pores, has significantly improved over the past decade \citep{shin_size-dependent_2016,battat_particle_2019,gupta_diffusiophoresis_2020,wilson_diffusiophoresis_2020,alessio_diffusiophoresis_2021,alessio_diffusioosmosis-driven_2022,shim_diffusiophoresis_2022,lee_role_2023,akdeniz_diffusiophoresis_2023}. However, in porous media flows, geometric disorder weakens the solute gradients \citep{koch_dispersion_1985,sahimi_flow_1993,dentz_mechanisms_2018,dentz_mixing_2023}. It is therefore not a priori clear if diffusiophoresis could play an important role in these complex environments. 

Recent experimental and theoretical works offer evidence for the role of diffusiophoresis in porous environments. In the absence of background flows, colloidal diffusiophoresis has been experimentally demonstrated and theoretically rationalized in fibrous porous media such as collagen gels and biofilms, where pore sizes are comparable to the colloid size \citep{doan_confinement-dependent_2021,sambamoorthy_diffusiophoresis_2023,somasundar_diffusiophoretic_2023,sambamoorthy_diffusiophoresis_2025}. In the presence of flows, microfluidic experiments and numerical simulations have illustrated how diffusiophoresis influences colloidal transport into and out of dead-end pores \citep{park_microfluidic_2021,jotkar_impact_2024}. More recently, we have shown how the interplay between flow disorder and solute gradients governs the transport of colloids in porous media, highlighting the role of diffusiophoresis in both preferential flow pathways and stagnant fluid pockets \citep{alipour_diffusiophoretic_2024}.

Despite these advances, it remains unclear whether diffusiophoretic removal of colloids from stagnant fluid pockets remains effective over larger scales, as solute fronts become increasingly diffuse, weakening spatial gradients and diffusiophoretic velocities. Here, we address this central question: does dispersion of the solute front weaken the phoretic effects? To address this question, here we combine numerical simulations and microfluidic experiments, demonstrating that solute front dispersion in disordered porous media does not diminish colloid removal efficiency. To rationalize this observation, we analyze diffusiophoretic colloid migration from idealized one-dimensional dead-end pores exposed to diffuse solute fronts, deriving an analytical model to describe the spatiotemporal evolution of solute and colloid fields. Our model shows excellent agreement with numerical simulations for linear and Error function solute profiles, showing that broadening of the solute front leads to a persistent spatial gradient beyond the solute diffusion timescale along the dead-end pore, leading to a more efficient and uniform extraction of colloids. Our work therefore points to the potential relevance of diffusiophoresis in realistic porous media scenarios.

\section{\label{sec:PM}Influence of solute dispersion on phoretic colloid removal from porous media}

\begin{figure*}
    \centering
    \includegraphics[width=\textwidth, keepaspectratio]{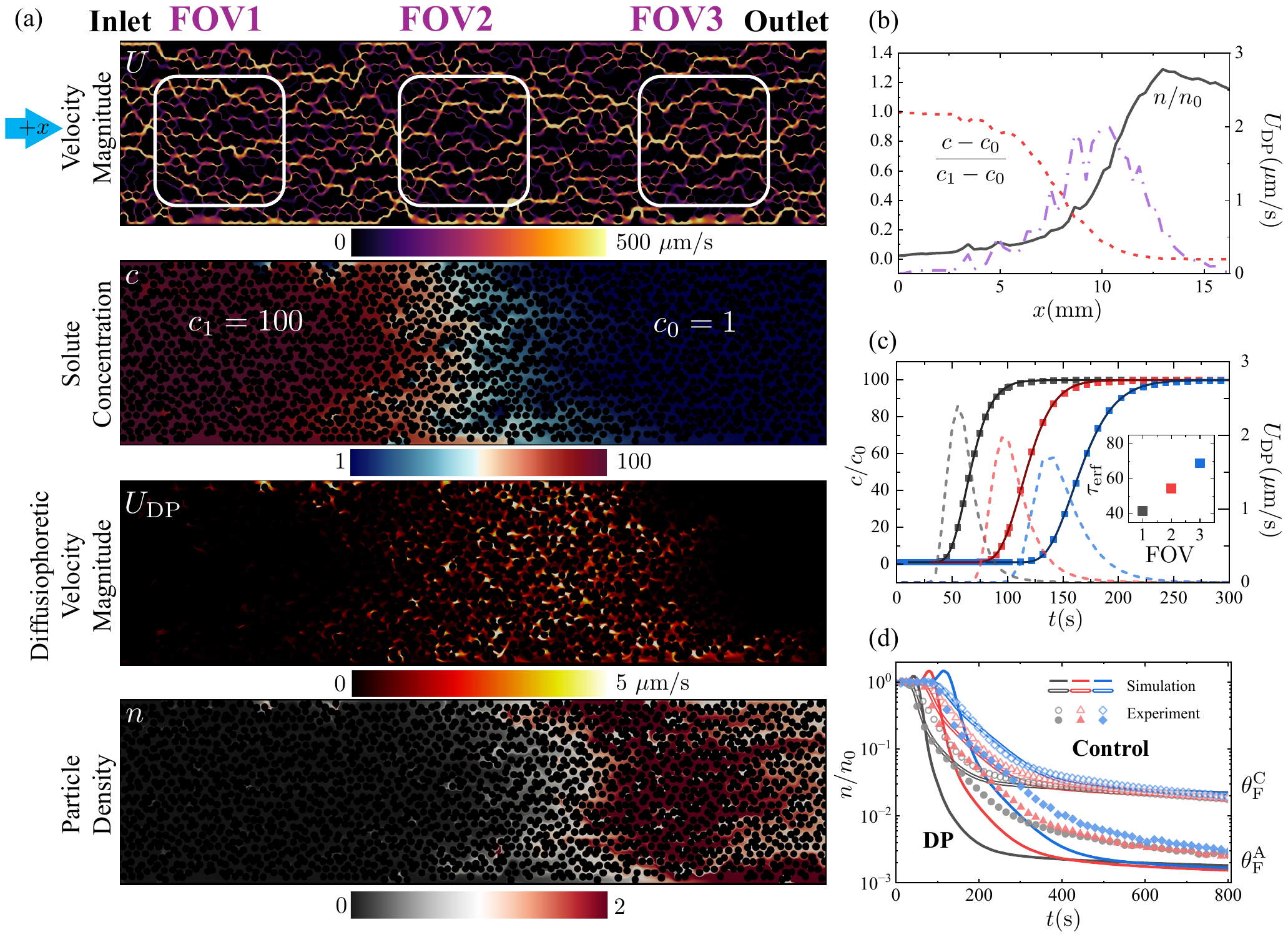}
    \captionsetup{width=\textwidth, justification=justified, singlelinecheck=false}  
    \caption{The dispersion of solute front does not weaken the diffusiophoretic removal of colloids. 
    (a–c) Numerical simulations; (d) Simulations and experiments.
    (a) We displace an aqueous solution of colloids with salt concentration $c_0$ with a colloid-free solution with salt concentration $c_1$ at constant flow rate. Medium disorder leads to the heterogeneous distribution of flow velocity magnitude, solute concentration, diffusiophoretic velocity magnitude and particle density. Three fields of view (FOVs) are highlighted by white boxes.
    (b) Cross-sectionally averaged profiles of the normalized solute concentration 
    (red dashed), particle density (black solid),
    and diffusiophoretic velocity magnitude  (purple dash-dotted), shown at the same time as panel (a).
    (c) Temporal evolution of the average solute concentration and diffusiophoretic velocity magnitude within the three FOVs. The solute profiles are fitted using the solution to the 1D advection–diffusion equation (Eq.~\eqref{eq:1Dsoln}). The inset presents the fitted characteristic solute transition time $\tau_\text{erf}$ for each FOV.
    (d) Temporal evolution of the average particle density across the three FOVs, in the presence (single lines), and absence of solute gradients (double solid lines). Symbols represent the corresponding experimental observations.}
    \label{fig:PM}
\end{figure*}

\noindent \textbf{Experimental setup:} We probe the transport of colloids in a quasi 2D microfluidic geometry with height $h \approx 50\ \mu\text{m}$, patterned with a disordered array of cylindrical obstacles with diameter $\approx 175\ \mu\text{m}$. The mean pore size is $\approx 60\ \mu\text{m}$ and the medium porosity is $\approx 0.36$. The medium is initially filled with an aqueous colloidal solution with density $n_0=1$ and solute concentration $c_0=1~\textrm{mM}$. We then inject a second aqueous solution with solute concentration $c_1=100~\textrm{mM}$ at constant flow rate to displace the colloids and monitor the removal process in different fields of view along the medium (Fig.~\ref{fig:PM} (a)). In the experiments, we use LiCl as the solute with diffusivity of \(D_\text{s} = 1.37 \times 10^{-9}\,\mathrm{m}^2/\mathrm{s}\), and 1 micron size colloids with diffusivity \(D_\text{p} = 4.37 \times 10^{-13}\,\mathrm{m}^2/\mathrm{s}\), and diffusiophoretic mobility of \(\Gamma_\text{p} \approx 0.7 \times 10^{-9}\,\mathrm{m}^2/\mathrm{s}\) \citep{shin_size-dependent_2016,velegol_origins_2016,gupta_diffusiophoresis_2020,alessio_diffusioosmosis-driven_2022,liu_diffusioosmotic_2025,alipour_diffusiophoretic_2024}. We note that the colloidal suspension is very dilute with a volume fraction of $\approx 0.1$ percent, leading to a packing fraction of $\approx 10^{-3}$, which together with the $1/r^3$ decay of hydrodynamic disturbance velocity for diffusiophoretic colloids implies that we can safely ignore particle-particle or particle-surface interactions \citep{shin_size-dependent_2016}. \\

\noindent \textbf{Numerical simulations:} We use OpenFOAM to numerically solve for the evolution of solute and colloid fields \citep{ault_diffusiophoresis_2018,ault_characterization_2019,liu_diffusioosmotic_2025,alipour_diffusiophoretic_2024}. We calculate the steady-state flow field using the \texttt{simpleFoam} solver, applying a pressure drop across the medium, which results in a mean velocity magnitude of $\langle U_{\text{mag}} \rangle = 184.3\ \mu\text{m/s}$. The geometric disorder of the medium leads to a heterogeneous velocity field, where most of the transport occurs through preferential flow pathways surrounding low permeability/stagnant fluid pockets (Fig.~\ref{fig:PM} (a)). With the steady-state flow field established, we then solve for the transient transport of solute, $c$, and particles, $n$, governed by the advection--diffusion equations:
\begin{equation}
\frac{\partial c}{\partial t} + \nabla \cdot (\mathbf{u} c) = D_{\text{s}} \nabla^2 c,
\label{eq:1}
\end{equation}
\begin{equation}
\frac{\partial n}{\partial t} + \nabla \cdot \left( (\mathbf{u} + \mathbf{u}_{\text{DP}}) n \right) = D_{\text{p}} \nabla^2 n,
\label{eq:2}
\end{equation}
where $\mathbf{u}$ is the background velocity, and $\mathbf{u}_{\text{DP}} = \Gamma_{\text{p}} \nabla \ln c$ represents the diffusiophoretic velocity. The recent work of \citet{jotkar_impact_2024} also studies the diffusiophoretic transport of colloids in porous media using these equations, reporting on the migration of colloids into/from dead-end pores due to solute gradients. \\ 

\noindent \textbf{Flow disorder enhances solute dispersion:} Flow disorder leads to a non-uniform solute front, enhancing the dispersion \citep{saffman_theory_1959,koch_dispersion_1985,koch_effect_1989,sahimi_flow_1993,dentz_mechanisms_2018,puyguiraud_pore-scale_2021,liu_scaling_2024,woods_dispersive_2025} and weakening the diffusiophoretic velocities (Fig.~\ref{fig:PM} (a)). The broadening of the solute profile leads to the persistence of solute gradients around the stagnant fluid pockets \citep{de_anna_chemotaxis_2021} as evidenced in the snapshot of the diffusiophoretic velocity field. The cross-sectionally averaged fields show the extent of dispersion along the medium (Fig.~\ref{fig:PM} (b)). We note that the peak of the diffusiophoretic velocity is ahead of the solute front due to the \(\nabla c / c\) functional form.

As the solute front travels along the medium, it becomes more dispersed, as demonstrated by the evolution of mean solute concentration in the three fields of view along the medium (Fig.~\ref{fig:PM} (c)). To characterize this dispersion, we fit the solute profiles using the analytical solution of the one-dimensional advection--diffusion equation:
\begin{equation}
c(t) = c_0 + \frac{c_1 - c_0}{2} \left( 1 + \text{Erf} \left( \frac{U_\text{x} t - x_{\text{FOV}}}{\sqrt{4D_{\text{eff}} t}} \right) \right),
\label{eq:1Dsoln}
\end{equation}
where $U_\text{x} = 141\ \mu \text{m/s}$ is the average velocity in the mean flow direction. Here, the effective distance from the inlet, $x_{\text{FOV}}$, and the effective dispersion coefficient, $D_{\text{eff}}$, are determined as fitting parameters. We can then define the characteristic solute transition time for the error-function profile as:
\begin{equation}
\tau_\text{erf} = \frac{2\sqrt{4D_{\text{eff}} (x_{\text{FOV}} / U_\text{x})}}{U_\text{x}}.
\end{equation}

We observe that $\tau_\text{erf}$ increases by $60\%$, from $41.7\ \text{s}$ in FOV1 to $68.6\ \text{s}$ in FOV3, over a distance of $10\ \text{mm}$, which corresponds to approximately $160$ pore sizes (inset of Figure \ref{fig:PM}(c)). These results highlight the significant dispersion of the solute front as it travels through the porous medium, further emphasizing the influence of the disordered structure on solute transport. In Fig.~\ref{fig:PM}(c), the evolution of the average magnitude of the diffusiophoretic velocity in each FOV is plotted with semi-transparent dashed lines. As expected, the increasingly dispersed solute front generates a weaker but more persistent diffusiophoretic velocity field in downstream FOVs compared to upstream regions.\\

\noindent \textbf{Solute dispersion does not weaken the phoretic removal:} To examine the influence of diffusiophoresis on particle transport, we monitor the temporal evolution of the average particle density, $n(t)/n_{0}$, for each FOV (Figure \ref{fig:PM}(d)). The results are represented by thick black, red, and blue lines. We further conduct simulations and experiments for the control case, where no solute gradients are present ($c_0=c_1$). The corresponding average particle density profile in this control case is plotted as double lines. In the control case, the average particle density decreases rapidly from its initial value of $1.0$ to approximately $n_{f}/n_0 \approx 0.015$, where $n_0$ and $n_f$ represent the initial and final particle density, respectively. This initial drop is primarily due to the displacement of particles in  main flow pathways. Subsequently, the particle density decreases more gradually, as diffusion becomes their only escape route from dead-end pores.

In the presence of solute gradients, the colloid removal becomes much more efficient as diffusiophoresis removes the colloids from dead-end pores and lead to their cross-streamline migration in the main flow pathways (Fig.~\ref{fig:PM}(d)) \citep{alipour_diffusiophoretic_2024}. We might expect the sharper solute front in the upstream region (FOV 1) to lead to a more efficient particle removal than the downstream region (FOV 3). Our observations, however, clearly show this not to be the case as the final particle fraction after the passage of solute front is the same in all three fields of view ($n_{f}/n_0 \approx 0.001$, as shown in figure \ref{fig:PM}(d)). Our numerical simulations agree reasonably well with our microfluidic experiments shown by the symbols in Fig.~\ref{fig:PM}(d). To gain insight into this surprising observation, we next focus on the idealized case of 1D dead-end pores, probing how solute dispersion influences the diffusiophoretic removal of colloids from stagnant fluid pockets.

\section{\label{sec:DEP_Linear}Diffusiophoretic removal of colloids from dead-end pores}

\begin{figure*}
    \centering
    \includegraphics[width=\textwidth, keepaspectratio]{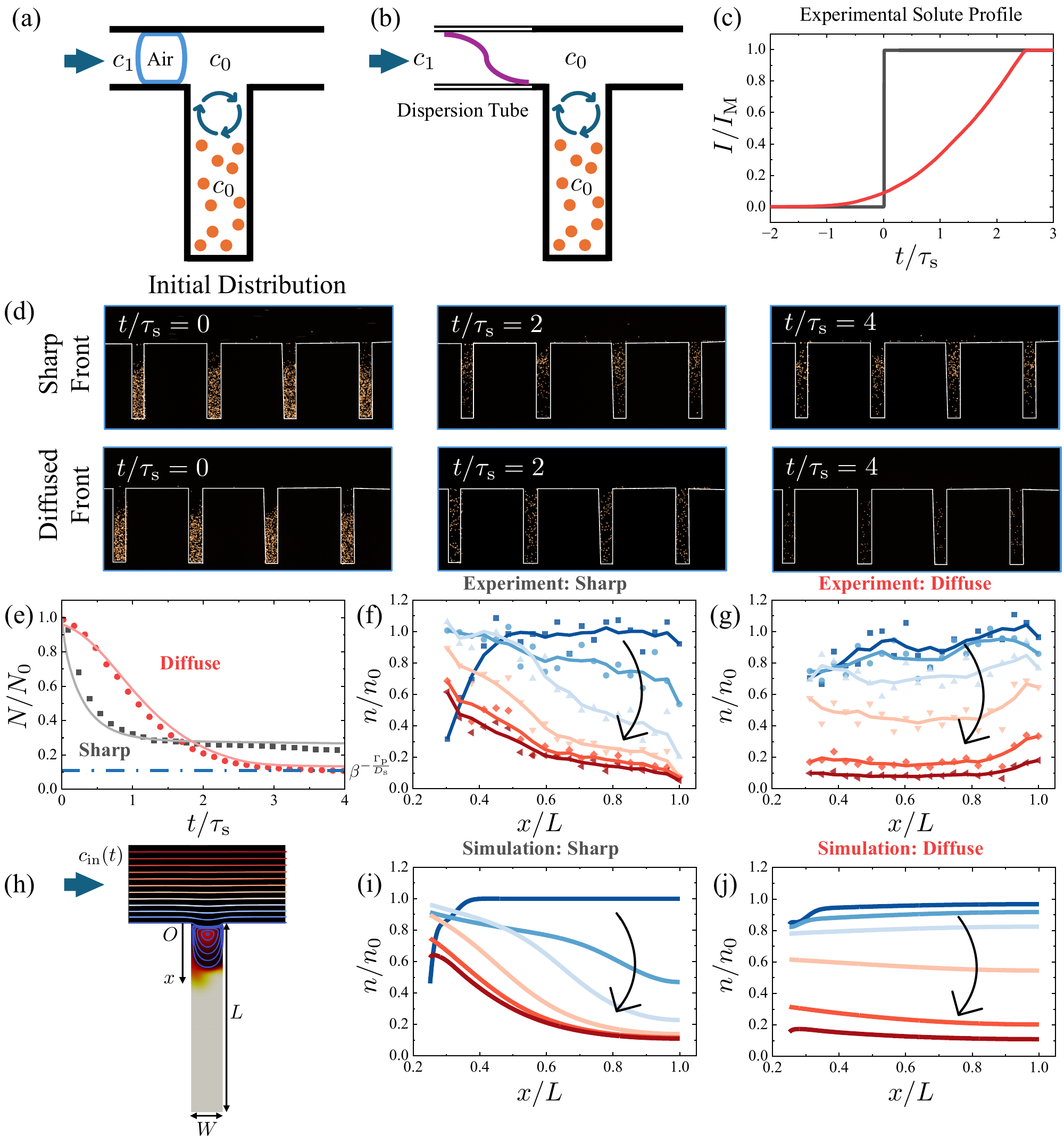}
    \captionsetup{width=\textwidth, justification=justified, singlelinecheck=false} 
    \caption{\label{fig:DEP_EXP}%
    The evolution of colloid density in dead-end pores exposed to (a) sharp, and (b) diffuse solute fronts. (c) The dispersion of solute front is represented by the evolution of the fluorescein light intensity at the inlet of the dead-end pores.
    (d) The evolution of colloid density exposed to sharp versus diffuse solute fronts. For both cases; snapshots at \(t/\tau_\text{s}=0,\,2,\,4\).
    (e) The evolution of fraction of residual particles over time showing the crossover at ``late times". Symbols: experiments; solid lines: 2D simulations. The blue dash–dot line indicates the analytical prediction of the final residual (equation~\eqref{eq:General_N_Ave_ND}), consistent with both.
    In the presence of sharp solute front, colloids migrate toward the inlet of the dead-end pore, leading to a non-uniform density profile (f). However, in the diffuse front case, colloidal density in the pore remains nearly uniform (g). For the sharp front, data points correspond to 
    \(t/\tau_\text{s}=0,\,0.1,\,0.2,\,0.5,\,1,\,2\), and for the diffuse front, we have 
    \(t/\tau_\text{s}=0,\,0.25,\,0.5,\,1,\,2,\,4\). Solid 
    lines represent smoothed density profiles obtained via a five-point 
    adjacent-averaging method.
    (h) Two-dimensional simulation domain. Colour map: particle density after the flushing stage; streamlines: background flow showing the primary recirculation in the pore.
    (i,j) Simulated particle density profiles for the sharp (i) and diffuse (j) fronts at the same sampling times as in (f,g).
    }
\end{figure*}

\subsection{Sharp vs diffuse solute fronts: experiments and 2D simulations}
\noindent \textbf{Experiments:} To probe the role of solute dispersion on the removal of colloids from dead-end pores, we conduct two sets of experiments with sharp and diffuse solute fronts. In both, we first fill the chip with an aqueous colloidal solution with solute concentration $c_0=1~\textrm{mM}$. We then displace this solution with a second colloid-free aqueous solution with salt concentration $c_1=100~\textrm{mM}$ (Fig.~\ref{fig:DEP_EXP} (a,b)). In each chip, six dead-end pores of length $600~\mu\textrm{m}$ and width $100~\mu\textrm{m}$ are connected to the main channel. To create a sharp step-like solute concentration gradient at the inlet of dead-end pores, we use an air bubble to separate the two solutions \citep{shin_size-dependent_2016,wilson_diffusiophoresis_2020,alessio_diffusioosmosis-driven_2022}. In the absence of this bubble, the interface between the two solutions becomes diffuse, leading to a slowly varying solute concentration profile. We added a trace amount of fluorescein dye to the colloid-free solution to probe the evolution of the solute concentration at the inlet of dead-end pores, demonstrating the distinction between the sharp and diffuse solute profiles (Fig.~\ref{fig:DEP_EXP} (c)). Here we use LiCl as the solute, matching the porous-media experiments above. We also repeated the experiments with NaCl, obtaining similar results (Appendix~A.3).

Surprisingly, displacement using a diffuse solute front leads to a more effective removal of the colloids (Fig.~\ref{fig:DEP_EXP} (d)). In the sharp front case, colloids migrate toward the inlet of dead-end pore, but not all can escape before the solute gradients vanish ($\tau_s=L^2/D_s \approx 250~\textrm{s}$). In the diffuse front case, however, the removal of particles is more uniform. This is evident in 1D colloid density profiles constructed using particle detection and averaged over the six dead-end pores  (Fig.~\ref{fig:DEP_EXP} (f,g)). The lines show an interpolation using the five-point adjacent-averaging method. 

The evolution of particle density in dead-end pores clearly shows that the sharp solute front is more effective in removing the colloids at ``early times". However, we observe a crossover at ``late times" to the diffuse solute front becoming the more efficient way to remove particles from dead-end pores (Fig.~\ref{fig:DEP_EXP} (e)). In the sharp solute front case, solute gradients disappear beyond $t=\tau_s$, and particle diffusivity is negligible over the timescales probed here since $L^2/D_p\gg L^2/D_s$. Colloids can migrate in dead-end pores as long as solute gradients persist. Therefore, we expect the persistence of solute gradients in the diffuse front case to lead to the crossover at late times.\\

\noindent \textbf{Numerical simulations:}
We solve the coupled solute and colloid transport in a two-dimensional (2D) dead-end pore attached to a straight main channel (Fig.~\ref{fig:DEP_EXP}(h)) using OpenFOAM. The pore length and width are \(L=600~\mu\mathrm{m}\) and \(W=100~\mu\mathrm{m}\), matching the experiments; the main channel has the same in-plane dimensions. A symmetry condition at the top wall halves the computational channel height relative to the experiments. The background flow is left-to-right with mean speed \(U_{\mathrm m}=2~\mathrm{mm\,s^{-1}}\) (measured experimentally). We initialize the particle number density uniformly at \(n_0\), apply a short flushing stage to clear high-velocity regions (the main channel and the first primary eddy in the pore), and then impose a time-dependent inlet solute history \(c_{\mathrm{in}}(t)\) at the left boundary. For a \emph{sharp} front we impose a step from \(c_0=1~\mathrm{mM}\) to \(c_1=100~\mathrm{mM}\). For a \emph{diffuse} front we use the fluorescein-based inlet profile measured in the experiments (Fig.~\ref{fig:DEP_EXP}(c)), but rescale its time axis by the factor \(\sqrt{D_{\text{f}}/D_{\text{s}}}\) to account for the different molecular diffusivities of salt and fluorescein, where \(D_\mathrm{s}=1.37\times10^{-9}~\mathrm{m^2\,s^{-1}}\) and \(D_\mathrm{f}\approx4.6\times10^{-10}~\mathrm{m^2\,s^{-1}}\).
This scaling follows from high-Péclet dispersion, for which an effective coefficient of the form $D_\text{eff}/D\sim \mathrm{Pe}^2$ yields a front width (and thus a transition time) \(\Delta t \sim (D_\text{eff})^{1/2}/U_{\mathrm m} \propto D^{-1/2}\); hence the fluorescein profile is compressed by \(\sqrt{D_{\text{f}}/D_{\text{s}}}\) to represent the salt front. 
Here we take the diffusiophoretic mobility to be constant, \(\Gamma_{\mathrm p}=0.7\times10^{-9}~\mathrm{m^2\,s^{-1}}\). We also repeat the simulations with the constant–surface–charge (CSC) model \citep{lee_role_2023}, which yields similar results (Appendix~A.4). Other transport parameters are as specified above; in particular, the particle diffusivity is \(D_{\mathrm p}=4.37\times10^{-13}~\mathrm{m^2\,s^{-1}}\).

As shown by the solid curves in Fig.~\ref{fig:DEP_EXP}(e), the simulations reproduce the experimental “crossover’’ in the residual fraction \(N/N_0\): sharp fronts remove particles more rapidly at early times, whereas diffuse fronts ultimately leave fewer particles. In both cases, the agreement with the measurements is close, and for the diffuse front the experiment and simulation converge to the same final residual once the solute gradient has vanished. The profiles in Fig.~\ref{fig:DEP_EXP}(i–j) further illustrate this behavior: the sharp front generates a non-uniform \(n/n_0\) with a peak near the inlet, while the diffuse front yields an almost uniform profile with the same final residual; both profiles agree reasonably with the experiments in Fig.~\ref{fig:DEP_EXP}(f-g).

\subsection{Sharp vs diffuse solute fronts: 1D model with a linear solute profile}

\noindent To elucidate why a diffuse solute front can enhance particle removal, we further simplify our model to a one-dimensional (1D) dead-end pore of length $L$, with the inlet located at $x = 0$ (Fig.~\ref{fig:DEP_Simu}(a)). The influence of flow within the dead-end pore is assumed negligible, which is a reasonable approximation for pores with high aspect ratio. The initial solute concentration and particle density are set to $c_0 = 1$ and $n_0 = 1$, respectively, with boundary conditions at the inlet $c_1 = c_\text{in}(t)$ and $n_1 = 0$. \\

For simplicity, we first impose a linearly varying inlet solute concentration profile $c_\text{in}(t)$ that increases from $c_0$ to $c_1$ over a duration $T$, as shown in figure \ref{fig:DEP_Simu}(a). The evolution of the solute concentration field is governed by the diffusion equation:
\begin{equation}
\frac{\partial c\left(x,t\right)}{\partial t} - D_{\text{s}}\frac{\partial^2 c\left(x,t\right)}{\partial x^2} = 0,
\label{eq:3}
\end{equation}
subject to the initial condition $c(x,0) = c_0$ and boundary conditions:
\begin{equation}
c\left(0,t\right) = 
\begin{cases} 
c_0 + \frac{c_1 - c_0}{T} \cdot t, & t \in [0, T), \\
c_1, & t \in [T, \infty),
\end{cases}
\qquad \frac{\partial c}{\partial x}(L,t) = 0,
\label{eq:linear_solute_profile}
\end{equation}
where $D_{\text{s}}$ is the solute diffusivity, and $L$ is the pore length. Defining the characteristic solute diffusion time as $\tau_{\text{s}} = L^2 / D_{\text{s}}$, we nondimensionalize the system using $\bar x = x / L$, $\bar t = t / \tau_{\text{s}}$, $\bar T = T / \tau_{\text{s}}$, and $\bar c = (c - c_0) / (c_1 - c_0)$. The resulting analytical solution for the nondimensional solute field is:
\begin{equation}
\bar c\left(\bar x, \bar t\right) =
\begin{cases} 
\sum\limits_{n=0}^\infty \frac{2}{\lambda_n^3 \bar T} e^{-\lambda_n^2 \bar t} \sin\left(\lambda_n \bar x\right) + \frac{1}{2\bar T} \left(\bar x^2 - 2\bar x\right) + \frac{\bar t}{\bar T}, & \bar t \in [0, \bar T], \\
-\sum\limits_{n=0}^\infty \frac{2}{\lambda_n^3 \bar T} \left(1 - e^{-\lambda_n^2 \bar T}\right) e^{-\lambda_n^2 (\bar t - \bar T)} \sin\left(\lambda_n \bar x\right) + 1, & \bar t \in (\bar T, \infty),
\end{cases}
\label{eq:solute_Analytical}
\end{equation}
where $\lambda_n = \frac{(2n+1)\pi}{2}$. 

Given that the particle diffusivity is significantly smaller than the solute diffusivity, we first assume particles to be non-diffusive. The trajectory $\bar x(\bar \xi_0, \bar t)$ of a particle starting at $\bar \xi_0 = \xi_0 / L$ is governed by:
\begin{equation}
\frac{\partial \bar x\left(\bar \xi_0, \bar t\right)}{\partial \bar t} = \bar U_{\text{DP}}\left(\bar x, \bar t\right) = \bar \Gamma_{\text{p}} \frac{\partial \ln\left(\bar c\left(\bar x, \bar t\right) + \frac{1}{\beta - 1}\right)}{\partial \bar x},
\label{eq:Udp_Expression1}
\end{equation}
where $\bar \Gamma_{\text{p}} = \Gamma_{\text{p}} / D_{\text{s}} \approx 0.7$, and $\beta = c_1 / c_0 = 100$.

The evolution of the solute, and diffusiophoretic velocity fields for cases with $\bar T = 0,\ 1,\ 10$ are shown in Fig.~\ref{fig:DEP_Simu}(c, d), obtained from Eq.~(\ref{eq:solute_Analytical}) and Eq.~(\ref{eq:Udp_Expression1}). For the $\bar T = 0$ case, L'Hospital's rule is applied to evaluate the $\bar T \rightarrow 0$ limit. In this case, the solute diffuses into the pore over timescale $\approx \tau_s$ leading to a rapid decay of the diffusiophoretic velocity. In contrast, for $\bar T \gg 1$, the normalized solute concentration profile approximately follows the form $\frac{1}{\bar T}(\bar x^2 - 2\bar x) + \frac{\bar t}{\bar T}$ as the exponential terms in Eq.~(\ref{eq:solute_Analytical}) decay rapidly (Fig.~\ref{fig:Gradual_Linear_Fields}(a)). This expression indicates that the solute concentration in the pore has a time-independent quadratic profile that increases at a constant rate to match the inlet boundary condition. This results in a weaker, but more persistent diffusiophporetic velocity compared to the sharp-gradient case (Fig.~\ref{fig:DEP_Simu}(b, d)).

Using Eqs.~(\ref{eq:solute_Analytical}) and (\ref{eq:Udp_Expression1}), we numerically determine the trajectories of non-diffusive particles, $\bar{x}(\bar{\xi}_0, \bar{t})$, for the three considered scenarios (Fig.~\ref{fig:DEP_Simu}(e)). Over time, the $x - \xi_0$ profiles shift downward, reflecting the diffusiophoretic migration of particles toward the inlet. The point where these profiles intersect the horizontal axis $x/L=0$ at any given moment defines the critical depth ($\xi_{0,c}$): particles originating above this depth ($\xi_0<\xi_{0,c}$) have exited the pore while those below ($\xi_0>\xi_{0,c}$) remain trapped. The red line in each panel denotes the fraction of residual particles when the solute gradient approaches zero. Increasing $T$ reduces the residual particle fraction, demonstrating that more dispersed solute fronts enhance particle extraction. Further, for larger $T$, the $x - \xi_0$ profiles become more linear, indicating a more uniform particle distribution.

To assess the effect of finite particle diffusivity, we solve the 1D advection--diffusion equation for particle density:
\begin{equation}
\frac{\partial n}{\partial \bar t} + \frac{\partial}{\partial \bar x}\left(\bar U_{\text{DP}} n\right) = \bar D_{\text{p}} \frac{\partial^2 n}{\partial \bar x^2},
\label{eq:ADE_Particle}
\end{equation}
where $\bar D_{\text{p}} = D_{\text{p}} / D_{\text{s}} \approx 3.2 \times 10^{-4}$ is the normalized particle diffusivity. The initial condition is $n(\bar x, 0) = n_0$, and the boundary conditions are $n(0, \bar t) = 0$ and $\frac{\partial n}{\partial \bar x}(1, \bar t) = 0$. Figure \ref{fig:DEP_Simu}(f) compares the non-diffusive particle density fields (solid black lines) obtained from the Lagrangian trajectories $n / n_0 = (\partial \bar{x} / \partial \bar{\xi}_0)^{-1}$ to the density fields obtained from  Eq.~(\ref{eq:ADE_Particle}), represented by the square symbols. The good agreement confirms that particle diffusivity has negligible influence on the timescales probed here, except in the vicinity of the pore inlet where a thin diffusion layer induces minor deviations.

\begin{center}
    \includegraphics[width=0.89\textwidth, keepaspectratio]{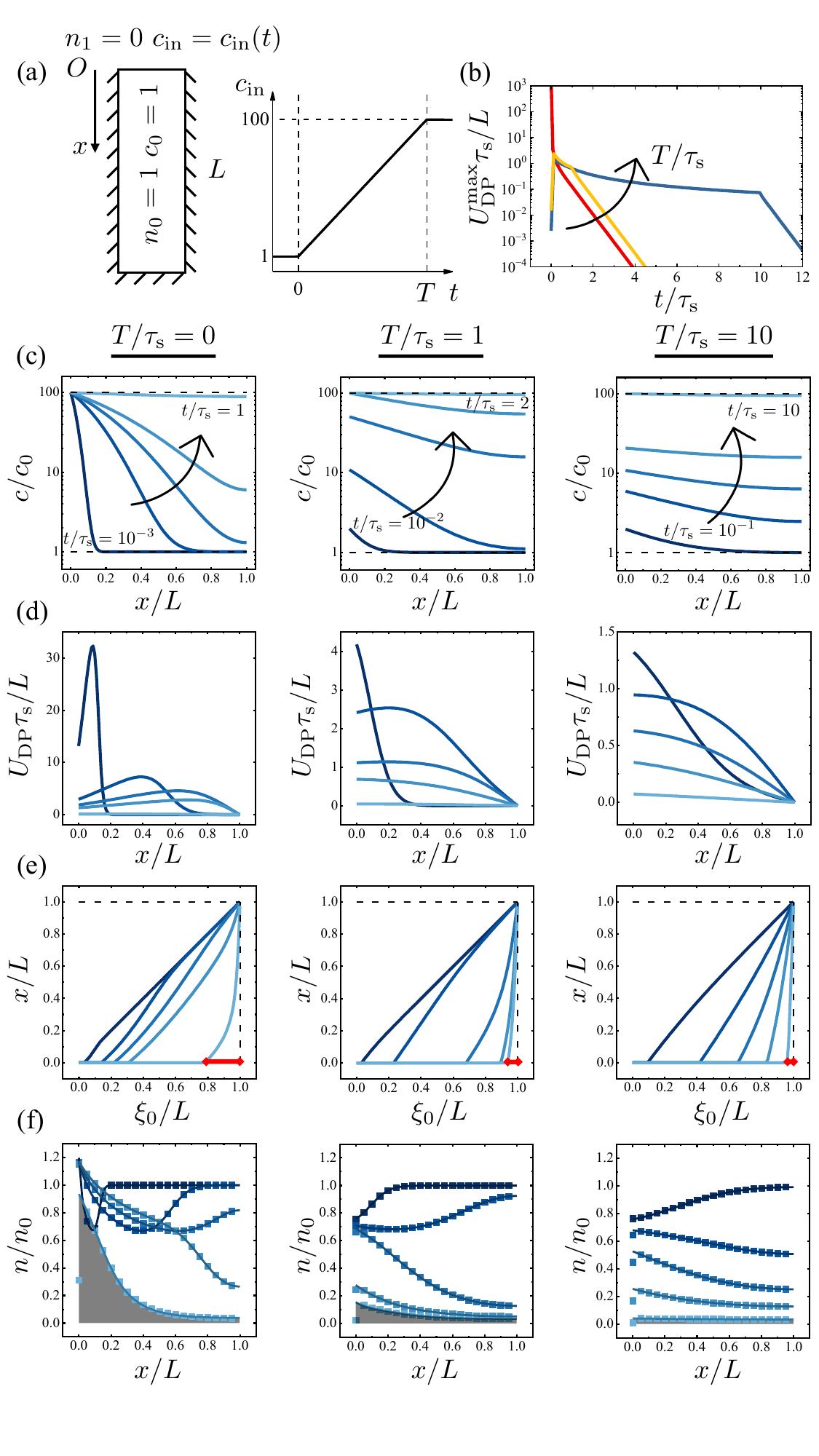}
\end{center}
\captionsetup{width=\textwidth, justification=justified, singlelinecheck=false}
\captionof{figure}{
Transport of solute and particles in a one-dimensional dead-end pore.
(a) Schematic of the setup: The initial solute concentration and particle density 
inside the pore are \(c_0 = 1\) and \(n_0 = 1\), respectively. The inlet boundary 
conditions are given by \(c_\text{in}(t)\) and \(n_1 = 0\). As shown on the right, 
\(c_\text{in}(t)\) increases linearly from 1 to 100 over the transition time \(T\) 
and remains at 100 thereafter.
(b) Maximum diffusiophoretic velocity over time for different normalized solute 
transition times \(T/\tau_\text{s} \in \{0,\,1,\,10\}\) (red: 0, yellow: 1, blue: 10).
(c, d) Evolution of the solute field (c) and diffusiophoretic velocity field (d) 
for \(T/\tau_\text{s} = 0,\,1,\,10\). Sampling times are 
\(t/\tau_\text{s} = \{0.001,\,0.02,\,0.05,\,0.1,\,1\}\) for \(T/\tau_\text{s} = 0\),
\(\{0.01,\,0.1,\,0.5,\,1,\,2\}\) for \(T/\tau_\text{s} = 1\), and
\(\{0.1,\,0.5,\,1,\,2,\,10\}\) for \(T/\tau_\text{s} = 10\).
(e) Particle trajectories \(x(t,\xi_0)\) as functions of both time \(t\) and 
initial position \(\xi_0\), using the same sampling times.
(f) Particle density fields: non-diffusive particles from particle tracking using Eq.~\eqref{eq:Udp_Expression1} (solid lines) and diffusive particles from the continuum model in Eq.~\eqref{eq:ADE_Particle} (symbols), evaluated at the same sampling times.}
\label{fig:DEP_Simu}

\vspace{1em}  

For a step-like solute boundary condition ($\bar{T} = 0$), the particle density profile is non-monotonic at early times, before the solute diffuses along the pore: particles in the middle region are rapidly drawn toward the inlet, whereas those near the bottom remain largely undisturbed. This results in a density peak near the inlet and a valley in the middle. As time progresses and the solute front dissipates, particles at the bottom also begin to move toward the inlet, ultimately producing a monotonic profile with a peak density of approximately $n / n_0 \approx 0.8$ at $\bar{t} = 1$.

In contrast, when the solute concentration changes gradually \((\bar{T} = 10)\), 
the particle density profile evolves smoothly, decreasing almost uniformly 
along the pore. By the time the diffusiophoretic velocity approaches zero, the 
profile is nearly flat, with a notably low density of \(n / n_0 \approx 0.03\). This  is consistent with our experimental observations (Fig.~\ref{fig:DEP_EXP}).
Because the particle diffusion time \(\tau_\text{p} = L^2 / D_\text{p}\) is much 
larger than the solute diffusion time \(\tau_\text{s}\), once the solute gradients 
(and thus the diffusiophoretic velocity) vanish, the particle distribution remains 
effectively unchanged. Hence, the gray shaded areas in 
Fig.~\ref{fig:DEP_Simu}(f), represent the unrecoverable particle fraction. \\

\begin{figure*}
    \centering
    \includegraphics[width=\textwidth, keepaspectratio]{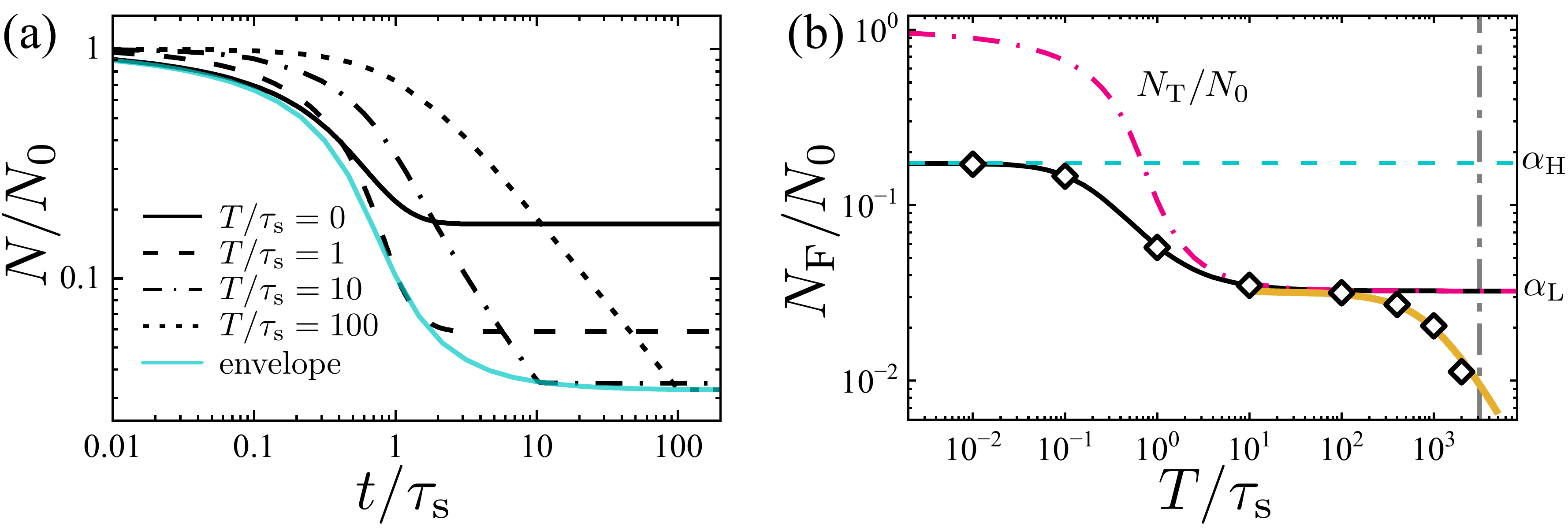}
    \captionsetup{width=\textwidth, justification=justified, singlelinecheck=false} 
    \caption{\label{fig:NF_T}
    Longer solute transition times slow the extraction of particles from dead-end pores but ultimately enhance their overall removal.
    (a) Temporal evolution of the fraction of residual particles in dead-end pores for different normalized solute transition times ($T/\tau_\text{s} = 0,\,1,\,10,\,100$), derived from the non-diffusive particle trajectories (Eq.~\eqref{eq:Udp_Expression1}), is shown by black lines. The cyan solid line is the envelope of these curves, indicating the minimal achievable residual particle fraction at any given time.
    (b) Final residual particle fraction (evaluated once the solute gradient has vanished) as a function of the solute transition time. Results are obtained by solving continuous equations (Eq.~\eqref{eq:ADE_Particle}) for diffusive particles (diamond symbols) and via particle tracking (Eq.~\eqref{eq:Udp_Expression1}) for non-diffusive particles (black solid lines). The magenta dash-dot line indicates the residual particle fraction at the moment when the inlet solute concentration reaches its final value, assuming particles are non-diffusive. The yellow solid line represents the prediction from our analytical model (Eq.~\eqref{eq:final_residual}) that accounts for both diffusiophoresis and particle diffusion.}
\end{figure*}

\noindent \textbf{Evolution of residual particle fraction:} To gain insight into the effect of solute concentration profile at the inlet, we use Eqs.(\ref{eq:solute_Analytical}) and (\ref{eq:Udp_Expression1}) to numerically solve for the evolution of residual particle fraction in the pores (Fig.~\ref{fig:NF_T}(a)). The residual fraction decreases initially before reaching a plateau at longer times when the solute gradients vanish. For larger $T/\tau_\text{s}$ values, the extraction process is slower, but more particles are ultimately removed. Consequently, at any given normalized time \(t/\tau_\text{s}\), there is an optimal \(T/\tau_\text{s}\) that maximizes particle extraction efficiency and minimizes the residual fraction. This optimum can be found numerically, and is represented by the cyan envelope in Fig.~\ref{fig:NF_T}(a).

The final residual particle fraction, \(\alpha\), decreases as $T$ increases, from $\alpha_{\text{H}}\approx0.17$ to a lower plateau of $\alpha_{\text{L}}\approx0.03$ as shown by the solid black line in Fig.~\ref{fig:NF_T}(b). This indicates that in the absence of particle diffusivity, a small fraction of particles remain unrecoverable regardless of how diffuse the solute front is. However, particle diffusion allows these particles to slowly diffuse out of the pores as shown by the diamond symbols obtained from the numerical solution of Eq.~\eqref{eq:ADE_Particle}. Note that the diamond symbols and the solid line match very well for times much smaller than particle diffusion timescale $T \ll \tau_\text{p} = L^2/D_\text{p}$. (gray vertical dash-dot line). 

\newpage

\section{\label{sec:DEP_Gradual}Analytical model for particle extraction with slowly varying solute profiles}

Consider a general, slowly varying inlet solute profile \(c_{\mathrm{in}}(\bar{t}) = C_{\mathrm{in}}(\bar{T}(\bar{t}))\), where the stretched time variable \(\bar T = \delta\,\bar t\) with $\delta \ll 1$ and the derivatives of \(C_{\mathrm{in}}\) with respect to \(\bar T\) are taken to be \(O(c_0)\). Thus, over the characteristic solute‐diffusion time \(\tau_{\mathrm{s}}\) (i.e. \(\Delta\bar t=O(1)\)), the relative change of the inlet concentration is only \(O(\delta)\), which defines the “slowly varying’’ limit. The solute field inside the pore evolves according to
\begin{equation}
\frac{\partial c(\bar{x},\bar{t})}{\partial \bar{t}} - \frac{\partial^2 c(\bar{x},\bar{t})}{\partial \bar{x}^2} = 0,
\label{eq:solute_general_eqn}
\end{equation}
with initial and boundary conditions
\begin{equation}
c(\bar{x},0) = c_\text{in}(0) = c_0,\qquad c(0,\bar{t}) = c_\text{in}(\bar{t}),\qquad \frac{\partial c}{\partial \bar{x}}(1,\bar{t}) = 0.
\label{eq:solute_IC_BCs}
\end{equation}
To homogenize the boundary conditions, we define
\begin{equation}
\phi(\bar{x}, \bar{t}) = c(\bar{x}, \bar{t}) - c_\text{in}(\bar{t}) - \tfrac{1}{2}(\bar{x}^2 - 2\bar{x})c_{\mathrm{in}}'(\bar{t})\,,
\end{equation}
where $c_{\mathrm{in}}' = \mathrm{d}c_{\mathrm{in}}/\mathrm{d}\bar{t}$. Substituting into Eq.~\eqref{eq:solute_general_eqn} yields
\begin{equation}
\frac{\partial \phi(\bar{x},\bar{t})}{\partial \bar{t}} - \frac{\partial^2 \phi(\bar{x},\bar{t})}{\partial \bar{x}^2} = -\tfrac{\delta^2}{2}(\bar{x}^2 - 2\bar{x})\,C_{\mathrm{in}}''(\bar{T}(\bar{t})),
\label{eq:solute_general_eqn_new}
\end{equation}
with the transformed initial and boundary conditions
\begin{equation}
\phi(\bar{x},0) = - \tfrac{\delta}{2}(\bar{x}^2 - 2\bar{x})\,C_{\mathrm{in}}'(0),\qquad \phi(0,\bar{t}) = 0,\qquad \frac{\partial \phi}{\partial \bar{x}}(1,\bar{t}) = 0.
\end{equation}
where \(C_{\mathrm{in}}'=\mathrm d C_{\mathrm{in}}/\mathrm d\bar T\) and \(C_{\mathrm{in}}''=\mathrm d^2 C_{\mathrm{in}}/\mathrm d\bar T^2\).

We expand \(\phi=\phi_0+\delta\,\phi_1+O(\delta^2)\). Since the source term on the right of \eqref{eq:solute_general_eqn_new} is \(O(\delta^2)\) and the initial condition is \(O(\delta)\), the leading problems are
\begin{equation}
    \phi_0\equiv0,\qquad
\frac{\partial \phi_1}{\partial \bar t}-\frac{\partial^2 \phi_1}{\partial \bar x^2}=0,\quad
\phi_1(\bar x,0)=-\tfrac12(\bar x^2-2\bar x)\,C_{\mathrm{in}}'(0),
\end{equation}
with \(\phi_1(0,\bar t)=0\) and \(\partial_{\bar x}\phi_1(1,\bar t)=0\). Using separation of variables with eigenfunctions \(\sin(\lambda_n\bar x)\), \(\lambda_n=\tfrac{(2n+1)\pi}{2}\) \((n=0,1,\dots)\), yields
\begin{equation}
\phi_1(\bar x,\bar t)=C_{\mathrm{in}}'(0)\sum_{n=0}^{\infty}\frac{2}{\lambda_n^{3}}\,e^{-\lambda_n^{2}\bar t}\,\sin(\lambda_n \bar x).
\end{equation}
Therefore, the solute field valid up to \(O(\delta)\), is
\begin{equation}
c(\bar x,\bar t)\approx
c_{\mathrm{in}}(\bar t)+\tfrac12(\bar x^2-2\bar x)\,c_{\mathrm{in}}'(\bar t)
+c_{\mathrm{in}}'(0)\sum_{n=0}^{\infty}\frac{2}{\lambda_n^{3}}\,e^{-\lambda_n^{2}\bar t}\,\sin(\lambda_n \bar x),
\label{eq:c_field_approximate}
\end{equation}
where \(\lambda_n=\tfrac{(2n+1)\pi}{2}\).

Since we focus on slowly varying solute profiles, the overall process duration is 
much longer than \(\tau_{\text{s}}\), ensuring that \(\bar{t} > 1\) for most of 
the extraction. Under these conditions, the exponentially decaying terms in 
Eq.~\eqref{eq:c_field_approximate} become negligible, effectively eliminating 
the influence of the initial condition. As a result, the solute field simplifies to
\begin{equation}
c(\bar{x}, \bar{t}) \approx c_\text{in}(\bar{t}) + \tfrac{1}{2}(\bar{x}^2 - 2\bar{x})\,c_\text{in}'(\bar{t}).
\label{eq:c_field_approximate1}
\end{equation}

To retain generality, we allow the (nondimensional) diffusiophoretic mobility to depend on concentration, \(\bar\Gamma_{\mathrm p}=\bar\Gamma_{\mathrm p}(c)\), which captures the dependence of the particle \(\zeta\)-potential and the Debye length on the local solute concentration \citep{prieve_motion_1984,lee_role_2023,gupta_diffusiophoresis_2020}. The diffusiophoretic velocity then reads
\begin{equation}
\bar U_{\mathrm{DP}}(\bar x,\bar t)
\;\simeq\;
\bar\Gamma_{\mathrm p}\!\left(c_{\mathrm{in}}(\bar t)\right)\,
\frac{(\bar x-1)\,c_{\mathrm{in}}'(\bar t)}{c_{\mathrm{in}}(\bar t)}\,,
\label{eq:DP_velocity}
\end{equation}
where we have ignored the slowly quadratic correction in Eq.~\eqref{eq:c_field_approximate1} compared with the leading term for slowly varying inlet profiles.
Therefore, the particle transport equation can be written as
\begin{equation}
\frac{\partial n}{\partial \bar{t}} + \frac{\partial}{\partial \bar{x}}\left[\bar{\Gamma}_\mathrm{p}(c_\text{in})\frac{(\bar{x}-1)c_\text{in}'}{c_\mathrm{in}}n\right] = \epsilon \frac{\partial^2 n}{\partial \bar{x}^2},
\label{eq:nfield_approx_eqn}
\end{equation}
where \(\epsilon = \bar{D}_\mathrm{p} = D_\mathrm{p}/D_\mathrm{s} \ll 1\), and subject to the initial and boundary conditions
\begin{equation}
n(\bar{x},0) = n_0,\quad n(0,\bar{t}) = 0,\quad \frac{\partial n}{\partial \bar{x}}(1,\bar{t}) = 0.
\end{equation}

We employ a singular perturbation analysis to solve Eq.~\eqref{eq:nfield_approx_eqn}. Far from the inlet, where diffusion is negligible, we set $\epsilon = 0$ and balance only the advective terms. By employing the method of characteristics, the outer solution, $n_{\mathrm{o}}$, is obtained
\begin{equation}
\frac{n_{\mathrm{o}}(\bar{x},\bar{t})}{n_0}=\frac{n_{\mathrm{o}}(\bar{t})}{n_0}
= \exp\!\left(-\int_{c_{\mathrm{in}}(0)}^{\,c_{\mathrm{in}}(\bar{t})} \bar{\Gamma}_{\mathrm p}(c)\,\frac{\mathrm{d}c}{c}\right).
\label{eq:outer_general}
\end{equation}
When \(\bar{\Gamma}_{\mathrm p}\) is constant, this reduces to the power law
\begin{equation}
\frac{n_{\mathrm{o}}(\bar{t})}{n_0}
= \left(\frac{c_{\mathrm{in}}(\bar{t})}{c_{\mathrm{in}}(0)}\right)^{-\bar{\Gamma}_{\mathrm p}}.
\label{eq:outer_constantGamma}
\end{equation}
The outer solution \(n_{\mathrm{o}}(\bar{t})\) is independent of \(\bar{x}\) and therefore cannot satisfy the inlet boundary condition \(n(0,\bar{t})=0\).

To resolve this mismatch, we consider a boundary layer near the inlet, introducing a stretched coordinate $\bar{y}=\bar{x}/\epsilon$. In the inner region, Eq.~\eqref{eq:nfield_approx_eqn} can be written as
\begin{equation}
\bar{\Gamma}_{\text{p}}\frac{c_\text{in}'(\bar{t})}{c_\text{in}(\bar{t})}\frac{\partial n_{\text{i}}}{\partial \bar{y}} + \frac{\partial^2 n_{\text{i}}}{\partial \bar{y}^2}
= \epsilon\left[\frac{\partial n_{\text{i}}}{\partial \bar{t}} + \bar{\Gamma}_{\text{p}}\frac{c_\text{in}'(\bar{t})}{c_\text{in}(\bar{t})}\left(n_{\text{i}} + \bar{y}\frac{\partial n_{\text{i}}}{\partial \bar{y}}\right)\right].
\end{equation}
Neglecting the $\mathcal{O}(\epsilon)$ terms and applying the boundary conditions $n_{\text{i}}(0,\bar{t})=0$ and $\lim_{\bar{y}\to\infty}n_{\text{i}}(\bar{y},\bar{t})=\lim_{\bar{x}\to 0}n_{\text{o}}(\bar{x},\bar{t})$, we find
\begin{equation}
\frac{n_{\text{i}}(\bar{y},\bar{t})}{n_0} = \frac{n_{\mathrm{o}}(\bar{t})}{n_0}
\left[1 - \exp\left(-\bar{\Gamma}_{\text{p}}\frac{c_\text{in}'(\bar{t})}{c_\text{in}(\bar{t})}\bar{y}\right)\right].
\end{equation}

Transforming back to the original coordinate, we have
\begin{equation}
\frac{n_{\text{i}}(\bar{x},\bar{t})}{n_0} = \frac{n_{\mathrm{o}}(\bar{t})}{n_0}
\left[1 - \exp\left(-\mathcal{K}(\bar t)\bar{x}\right)\right],
\label{eq:innerSolution}
\end{equation}
where for simplicity we define the diffusion layer thickness, \(\mathcal{K}^{-1}(\bar t) \equiv \left(\frac{\bar{\Gamma}_{\mathrm p}}{\bar{D}_{\mathrm p}}
\frac{ c_\mathrm{in}'(\bar t)}{ c_\mathrm{in}(\bar t)}\right)^{-1}\).

To construct the composite solution that satisfies both the inlet boundary condition and the behavior away from the inlet, we combine the inner and outer solutions as
\begin{equation}
\frac{n(\bar{x},\bar{t})}{n_0} = \underbrace{\frac{n_{\mathrm{o}}(\bar{t})}{n_0}}_{\frac{n_{\text{o}}}{n_0}}
+ \underbrace{\frac{n_{\mathrm{o}}(\bar{t})}{n_0}\left[1 - \exp\left(-\mathcal{K}(\bar t)\bar{x}\right)\right]}_{\frac{n_{\text{i}}}{n_0}}
- \underbrace{\frac{n_{\mathrm{o}}(\bar{t})}{n_0}}_{\lim_{\bar{x}\to 0}\frac{n_{\text{o}}}{n_0}},
\end{equation}
which simplifies to
\begin{equation}
\frac{n(\bar{x},\bar{t})}{n_0} = \exp\!\left(-\int_{c_{\mathrm{in}}(0)}^{\,c_{\mathrm{in}}(\bar{t})} \bar{\Gamma}_{\mathrm p}(c)\,\frac{\mathrm{d}c}{c}\right)
\left[1 - \exp\left(-\mathcal{K}(\bar t)\bar{x}\right)\right],
\label{eq:nField_composite}
\end{equation}
that is just the inner solution as the outer solution is constant.

To ensure the validity of this perturbation solution, we substitute it back into Eq.~\eqref{eq:nfield_approx_eqn} and verify that the diffusion term remains negligible compared to the advective term in the outer region. This leads to the thin diffusion layer condition
\begin{equation}
\lvert\mathcal{K}^{-1}(\bar t)\rvert = \bigg\lvert\left(\frac{\bar{\Gamma}_{\text{p}}}{\bar{D}_{\text{p}}}\frac{ c_\text{in}'(\bar{t}) }{ c_\text{in}(\bar{t}) }\right)^{-1}\bigg\rvert \ll 1.
\label{eq:condition2}
\end{equation}

Integrating Eq.~\eqref{eq:nField_composite} over $\bar{x}$ from 0 to 1 yields the average particle density:
\begin{equation}
\frac{n_{\text{ave}}(\bar{t})}{n_0} = \exp\!\left(-\int_{c_{\mathrm{in}}(0)}^{\,c_{\mathrm{in}}(\bar{t})} \bar{\Gamma}_{\mathrm p}(c)\,\frac{\mathrm{d}c}{c}\right)
\left[1 - \frac{1}{\mathcal{K}(\bar t)}
\left(1 - \exp\left(-\mathcal{K}(\bar t)\right)\right)\right].
\label{eq:General_N_Ave}
\end{equation}

In the non-diffusive limit, $D_{\text{p}}/\Gamma_{\text{p}} \rightarrow 0$, and with constant diffusiophoretic mobility, the final residual particle fraction simplifies to
\begin{equation}
\frac{N_{\text{F}}}{N_0} = \lim_{\bar{D}_{\text{p}} \rightarrow 0}\frac{n_{\text{ave}}(\infty)}{n_{0}} = \left(\frac{c_\text{in}(\infty)}{c_\text{in}(0)}\right)^{-\bar{\Gamma}_{\text{p}}} = \beta^{-\Gamma_{\text{p}}/D_{\text{s}}},
\label{eq:General_N_Ave_ND}
\end{equation}
where $\beta = c_\text{in}(\infty)/c_\text{in}(0)$. Here, \(N_0\) and \(N_\text{F}\) denote the initial and final total number of particles in the dead-end pore, while \(n_0\) denotes the initial particle number density in the pore. Notably, this expression is independent of the solute transition time and depends solely on the final solute contrast and the ratio $\Gamma_{\text{p}}/D_{\text{s}}$. This result is fully consistent with that of \citet{migacz_enhanced_2024}, which focused on non-diffusive 
particles and linear solute profiles from a Lagrangian point of view. In 
contrast, we adopt an Eulerian perspective by solving the continuous advection--diffusion 
equation for the particle density field, thereby incorporating finite particle 
diffusivity into the model. This approach captures the diffusion-driven 
corrections that become important for large \(T/\tau_{\text{s}}\). Moreover, our framework is applicable to a wide range of slowly varying inlet solute concentration
profiles, including those of the error-function type (Appendix~A.2).

\subsection{\label{subsec:Linear_Gradual}Linear solute profile under slowly varying conditions with constant \(\bar{\Gamma}_\text{p}\)}

To apply the above framework to the previously examined linear solute profile, we must first ensure that the inlet solute concentration is slowly varying and the thin diffusion layer condition~(\ref{eq:condition2}) is satisfied, which imply
\begin{equation}
\bar{T} \gg 1 \ \text{and}\ \lvert \mathcal{K}_\text{linear}(\bar t) \rvert \gg 1,\ \bar t\in [0,\bar T],
\label{eq:Linear_Case_Condition}
\end{equation}
where $\beta = c_1/c_0$ and \(\mathcal{K}_\text{linear}(\bar t) = \frac{\bar \Gamma_{\text{p}}}{\bar D_{\text{p}}} \frac{1}{\bar t + \frac{\bar T}{\beta-1}}\). Under these constraints, the particle density field and the average particle density simplify to:
\begin{equation}
\frac{n(\bar x,\bar t)}{n_{0}} = \left(1+\frac{\beta-1}{\bar T}\bar t\right)^{-\bar \Gamma_{\text{p}}}\left[1-\exp\left(-\mathcal{K}_\text{linear}(\bar t)\bar x\right)\right],\ \bar t\in [0,\bar T],
\label{eq:nField_Linear_Approx}
\end{equation}
and
\begin{equation}
\frac{n_{\text{ave}}(\bar t)}{n_{0}} = \left(1+\frac{\beta-1}{\bar T}\bar t\right)^{-\bar \Gamma_{\text{p}}}\left[1-\frac{1}{\mathcal{K}_\text{linear}(\bar t)}\left(1-\exp{\left(-\mathcal{K}_\text{linear}(\bar t)\right)}\right)\right],\ \bar t\in [0,\bar T].
\label{eq:n_ave_Linear_Approx}
\end{equation}

The above expressions apply only for \(\bar{t} < \bar{T}\). However, when \(\bar{T} \gg 1\), diffusiophoretic effects after \(\bar{t} = \bar{T}\) are negligible. As shown in figure~\ref{fig:NF_T}, for \(\bar{T} \gg 1\), the magenta dash-dot line, which indicates the residual particle fraction at \(\bar{t} = \bar{T}\), converges to the black solid line, representing the final residual particle fraction derived under the non-diffusive assumption. Therefore, for \(\bar{T} \gg 1\), we can approximate the final residual particle fraction by
\begin{equation}
\frac{N_\text{F}}{N_{0}} \approx \frac{n_{\text{ave}}(\bar T)}{n_{0}} = \beta^{-\bar \Gamma_{\text{p}}}\left[1-\frac{\bar D_{\text{p}}}{\bar \Gamma_{\text{p}}}\frac{\beta \bar T}{\beta -1}\left(1-\exp\left(-\frac{\bar \Gamma_{\text{p}}}{\bar D_{\text{p}}}\frac{\beta -1}{\beta \bar T}\right)\right)\right].
\label{eq:final_residual}
\end{equation}

Figure~\ref{fig:NF_T}(b) compares this prediction (yellow solid line) against the simulation results (diamonds). As expected, when $\bar T$ falls within $[10, \frac{\Gamma_{\text{p}}}{D_{\text{p}}}\cdot\frac{\beta-1}{\beta}]$, where condition~(\ref{eq:Linear_Case_Condition}) remains valid for most of the extraction process, the theoretical prediction closely matches the numerical simulations. The validity of the solute-field expression (Eq.~\eqref{eq:c_field_approximate1}) and the particle-field expression (Eq.~\eqref{eq:nField_composite}) is also examined (Appendix~A.1).

\begin{figure*}
    \centering
    \includegraphics[width=\textwidth, keepaspectratio]{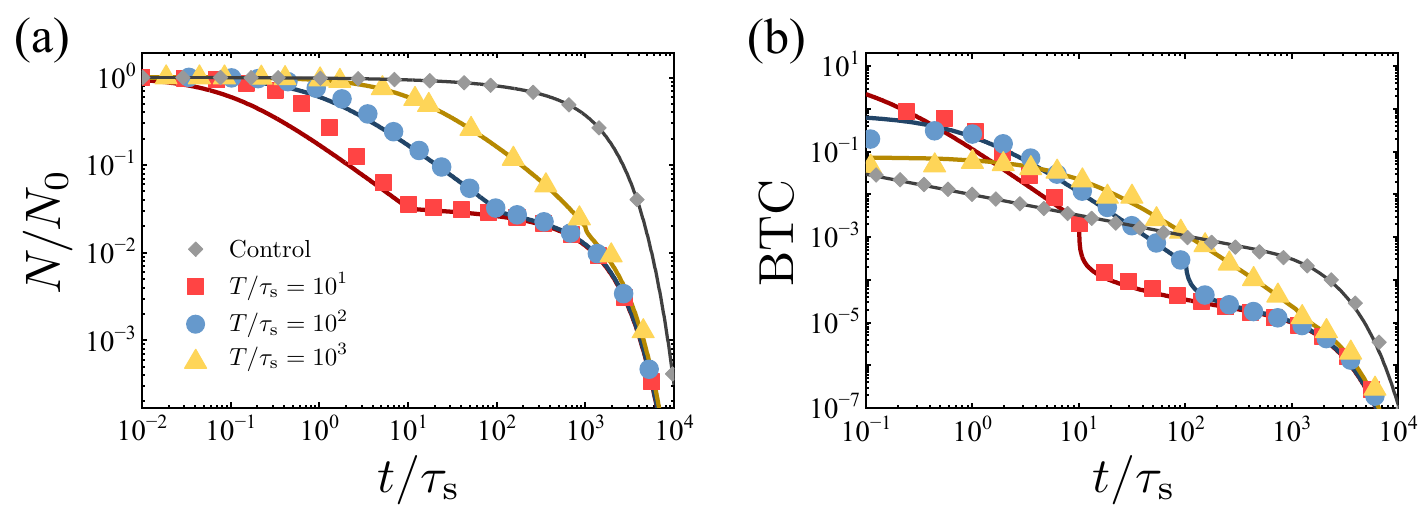}
    \captionsetup{width=\textwidth, justification=justified, singlelinecheck=false} 
    \caption{\label{fig:Gradual_Linear_Nt_BTC} 
    Analytical and numerical results for particle transport in a dead-end pore with large normalized solute transition times, $T/\tau_\text{s}$. (a, b) Residual particle fraction (a) and breakthrough curves (b) for $T/\tau_\text{s} = 10^1, 10^2$, and $10^3$, compared to a control case without solute gradients. Symbols indicate numerical results, and solid lines show analytical predictions, demonstrating good agreement across a range of solute transition times.}
\end{figure*}

Once the solute gradients vanish ($\bar{t} > \bar{T}$), particle transport is governed solely by diffusion. For a pure diffusion process starting from a uniform particle distribution $n_0$, the evolution of the average particle density is given by
\begin{equation}
\frac{n_{\text{ave}}^{\text{C}}(\bar{t})}{n_0} = \sum_{m=0}^{\infty} \frac{2}{\lambda_m^2}\exp\left(-\lambda_m^2 \bar{D}_{\text{p}}\bar{t}\right),
\label{eq:pure_diffusion}
\end{equation}
where $\lambda_m = \tfrac{(2m+1)\pi}{2}$. Under conditions of large solute transition times, the particle distribution remains nearly uniform; therefore, its concentration at $\bar{t} = \bar{T}$ can be used as the initial condition for the subsequent pure diffusion. By combining Eq.~\eqref{eq:pure_diffusion} with Eq.~\eqref{eq:n_ave_Linear_Approx}, we derive an expression valid over the entire process:
\begin{equation}
\frac{n_{\text{ave}}(\bar{t})}{n_0} = 
\begin{cases} 
\displaystyle\left(1+\frac{\beta-1}{\bar{T}}\bar{t}\right)^{-\bar{\Gamma}_{\text{p}}}\left[1-\frac{1}{\mathcal{K}_\text{linear}(\bar t)}\left(1-\exp\left(-\mathcal{K}_\text{linear}(\bar t)\right)\right)\right], & \bar{t}\in[0,\bar{T}], \\[8pt]
\displaystyle\frac{n_{\text{ave}}(\bar{T})}{n_0}\sum_{m=0}^{\infty}\frac{2}{\lambda_m^2}\exp\left(-\lambda_m^2\bar{D}_{\text{p}}(\bar{t}-\bar{T})\right), & \bar{t}>\bar{T}.
\end{cases}
\label{eq:all_time_linear}
\end{equation}

This composite solution characterizes the entire particle extraction process, encompassing the initial diffusiophoretic-driven phase ($\bar{t} \leq \bar{T}$) and the subsequent diffusion-only phase ($\bar{t} > \bar{T}$). Using this equation, we can also derive the particle escape time probability density function, or equivalently, the breakthrough curve (BTC), defined as
\begin{equation}
    p(\bar t) = - \frac{1}{n_0}\frac{d n_{\text{ave}}(\bar t)}{d\bar t}.
\end{equation}

Figure~\ref{fig:Gradual_Linear_Nt_BTC}(a) and  (b) respectively show the residual particle fraction and BTCs for linear solute profiles at various $\bar{T}$ values, along with a control case for comparison. Symbols represent numerical solutions of the advection–diffusion equations, and solid lines are our theoretical predictions. For the cases with $\bar{T}=10$ and $\bar{T}=100$, the BTCs display a distinct transition at $\bar{t}=\bar{T}$, reflecting the shift in escape mechanisms from combined diffusiophoretic and diffusive transport to pure diffusion after the solute gradient dissipates. 

\section[]{Solute concentration dependent diffusiophoretic mobility ($\Gamma_p(c)$)}\label{sec:cEffect}

So far, we have assumed the diffusiophoretic mobility to be constant. In practice, the local solute concentration modulates the particle zeta potential \(\zeta\) and the Debye length, and thus the mobility. To test the robustness of our conclusions, we retain the slowly varying, linear inlet profile but allow the nondimensional mobility \(\bar{\Gamma}_{\mathrm p}(c)\equiv \Gamma_{\mathrm p}(c)/D_{\mathrm s}\) to depend on \(c\).
The diffusiophoretic mobility is expressed as \citep{prieve_motion_1984,lee_role_2023}
\begin{equation}
\Gamma_{\mathrm p}
= \frac{\varepsilon}{\mu}\!\left(\frac{k_{B}T}{Z e}\right)^{2}
\!\left\{
\frac{Z e\,\zeta\,b}{k_{B}T}\,g(\lambda)
+ 4\,\ln\!\Big[\cosh\!\Big(\tfrac{Z e\,\zeta}{4 k_{B}T}\Big)\Big]\,h(\lambda)
\right\},
\label{eq:Gp_general}
\end{equation}
with
\begin{subequations}\label{eq:gh_defs}
\begin{align}
g(\lambda)
&= \left[
1 - \frac{\lambda k_{B}T}{2 Z e\,\zeta}
\!\left(
F_{1}(\bar{\zeta})
+ \frac{\varepsilon}{2\mu D_{\mathrm s}}\!\left(\frac{k_{B}T}{Z e}\right)^{2}
\big(b F_{4}(\bar{\zeta})+F_{5}(\bar{\zeta})\big)
\right)\!
\right]^{-1}, 
\\[4pt]
h(\lambda)
&= \left[
1 - \frac{\lambda}{8\,\ln\!\Big(\cosh\!\big(\tfrac{Z e\,\zeta}{4 k_{B}T}\big)\Big)}
\!\left(
F_{0}(\bar{\zeta})
+ \frac{\varepsilon}{2\mu D_{\mathrm s}}\!\left(\frac{k_{B}T}{Z e}\right)^{2}
\big(F_{2}(\bar{\zeta})+b F_{3}(\bar{\zeta})\big)
\right)\!
\right]^{-1}.
\end{align}
\end{subequations}

Here, \(\varepsilon\) is the medium permittivity, \(\mu\) the dynamic viscosity, \(k_B\) Boltzmann’s constant, \(T\) temperature, \(Z\) the electrolyte valence, \(e\) the elementary charge, and \(\zeta\) the zeta potential; \(\bar{\zeta}=\zeta/\!\big(k_B T/(Z e)\big)\) is its nondimensional form. The coefficient \(b=(D_{+}-D_{-})/(D_{+}+D_{-})\) (for a \(Z{:}Z\) electrolyte) involves the ionic diffusivities \(D_{\pm}\).
The size–dependent functions \(g(\lambda)\) and \(h(\lambda)\) depend on \(\lambda=(\kappa a)^{-1}\), the ratio of Debye length to particle radius \(a\), with \(\kappa^{-1}=\sqrt{\frac{\varepsilon k_B N_A T}{2 Z^{2} e^{2} c}}\), where \(N_A\) is Avogadro’s number and \(c\) the molar ion concentration \citep{prieve_motion_1984,lee_role_2023}. Following \citet{lee_role_2023}, we evaluate the auxiliary functions \(F_n(\bar{\zeta})\) using fitted forms \(F_n(\bar{\zeta})\approx -A_n \exp(\bar{\zeta}/T_n)\) with \(A_n=\{4.53,\,5.25,\,4.30,\,3.35,\,6.77,\,9.26\}\) and \(T_n=\{1.37,\,1.41,\,1.11,\,1.08,\,1.17,\,1.23\}\) for \(n=0,1,\ldots,5\).

To close the model, we relate the \(\zeta\) potential to the local salt concentration. Under a constant–surface–charge (CSC) approximation \citep{lee_role_2023}, \(\zeta\) varies approximately logarithmically with concentration. For a \(Z{:}Z\) electrolyte we adopt the empirical fit \citep{kirby_zeta_2004,kirby_zeta_2004-1}
\begin{equation}
  \zeta(c) \;=\; a_{0} \;+\; a_{1}\,\log_{10}\!\Big(\frac{Z^{2}\,c}{c_{0}}\Big),
  \label{eq:zeta_vs_C}
\end{equation}
where \(c_{0}=1~\mathrm{M}\) is a reference concentration, \(a_{0}\) (mV) is an offset, and \(a_{1}\) (mV/decade) is the slope. For latex particles in NaCl (\(Z=1\)) solutions, representative values are \(a_{0}\approx 0~\mathrm{mV}\) and \(a_{1}\approx -43.7~\mathrm{mV/decade}\) \citep{lee_role_2023,staffeld_diffusion-induced_1989}.

Substituting \eqref{eq:zeta_vs_C} together with
\(\lambda(c)=\big[\kappa(c)a\big]^{-1}\) (with \(\kappa\) the inverse Debye length) into \eqref{eq:Gp_general}–\eqref{eq:gh_defs} yields
\begin{equation}
  \bar{\Gamma}_{\mathrm p}(c)
  \;=\; \frac{1}{D_{\mathrm s}}\,
  \Gamma_{\mathrm p}\!\big(\zeta(c),\,\lambda(c)\big),
  \label{eq:Gp(c)}
\end{equation}
which we use below to quantify how local concentration modulates the particle mobility.

\begin{figure*}
    \centering
    \includegraphics[width=\textwidth, keepaspectratio]{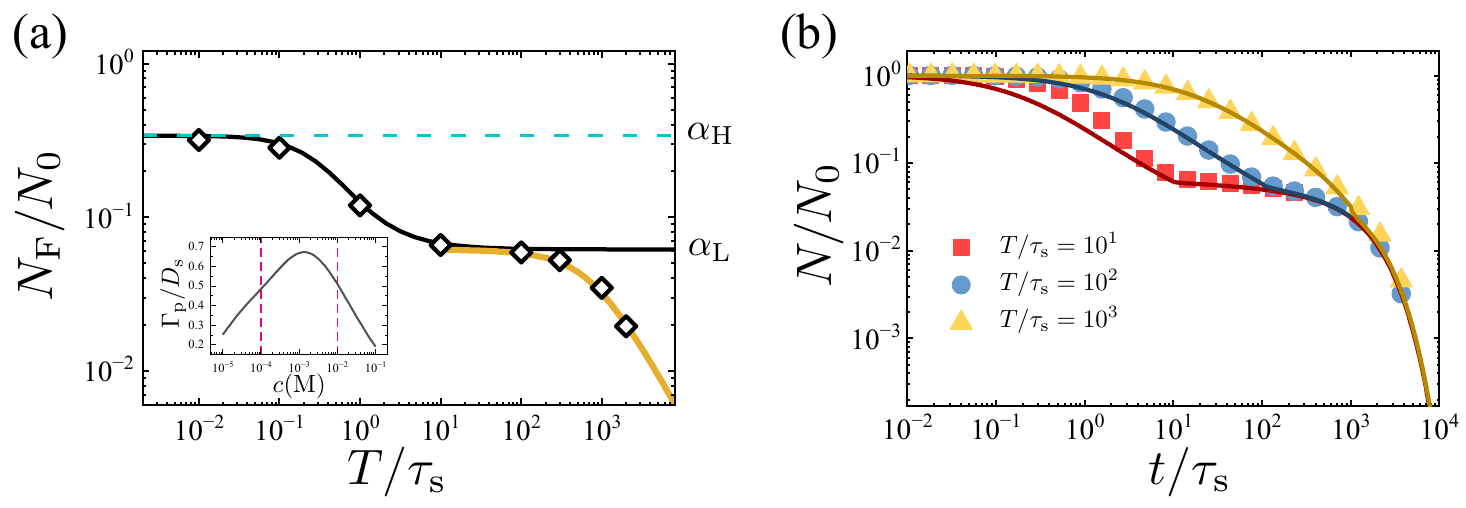}
    \captionsetup{width=\textwidth, justification=justified, singlelinecheck=false}
    \caption{\label{fig:CSC}
    Our analytical and numerical predictions using a solute concentration-dependent diffusiophoretic mobility $\Gamma_\mathrm{p}(c)$ mirrors those obtained earlier using constant diffusiophoretic mobility: longer solute–transition times enhance particle extraction.
    (a) Final residual fraction \(N_\mathrm{F}/N_0\) versus \(T/\tau_\mathrm{s}\).
    Diamonds: diffusive model from the continuum solution of \eqref{eq:ADE_Particle}.
    Black solid line: non-diffusive particle tracking using \eqref{eq:Udp_Expression1}.
    Yellow line: analytical prediction \eqref{eq:final_residual_CSC} that accounts for both diffusiophoresis and particle diffusion. Inset: predicted dependence of the normalized diffusiophoretic mobility \(\Gamma_\mathrm{p}/D_\mathrm{s}\) on bulk NaCl concentration \(c\) for latex particles by equation \eqref{eq:Gp(c)}; dashed magenta lines delimit the concentration range used in our simulations.
    (b) Temporal evolution of the residual fraction for \(T/\tau_\mathrm{s}=10^{1},10^{2},10^{3}\).
    Symbols: numerical results; solid lines: analytical predictions from \eqref{eq:n_ave_Linear_Approx_CSC}.
    In both panels the agreement is good across a wide range of solute–transition times.
    }
\end{figure*}

For a linear inlet ramp from \(0.1~\mathrm{mM}\) to \(10~\mathrm{mM}\), we numerically solve equations~\eqref{eq:Udp_Expression1} and \eqref{eq:ADE_Particle} with the concentration–dependent mobility \(\Gamma_\text{p}(c)\), and obtain the final residual particle fraction \(N_\text{F}/N_0\) as a function of the nondimensional solute transition time \(T/\tau_\text{s}\) (figure~\ref{fig:CSC}\,(a)). The trend mirrors the constant–\(\Gamma_\text{p}\) case: as \(T/\tau_\text{s}\) increases from \(0\) to \(\infty\), the non-diffusive model predicts a decrease in the final residual from \(\alpha_\text{H}=0.33\) to \(\alpha_\text{L}=0.06\). Including particle diffusion further reduces the residual at larger \(T\), as indicated by the symbols in figure~\ref{fig:CSC}\,(a).

The analytical framework for slowly varying inlet profiles also applies when \(\Gamma_\text{p}\) depends on \(c\). Following the procedure in subsection~\ref{subsec:Linear_Gradual}, but now with \(\Gamma_\text{p}=\Gamma_\text{p}(c)\), the particle density field and its cross-sectional average are obtained by combining \eqref{eq:Gp(c)}, \eqref{eq:nField_composite}, and \eqref{eq:General_N_Ave}:
\begin{equation}
\frac{n(\bar x,\bar t)}{n_{0}} =
\exp\!\left(-\int_{c_{\mathrm{in}}(0)}^{\,c_{\mathrm{in}}(\bar{t})}
\bar{\Gamma}_{\mathrm p}(c)\,\frac{\mathrm{d}c}{c}\right)
\left[1-\exp\!\left(-\mathcal{K}_\text{linear}(\bar t)\,\bar x\right)\right],
\quad \bar t\in [0,\bar T],
\label{eq:nField_Linear_Approx_CSC}
\end{equation}
\begin{equation}
\frac{n_{\text{ave}}(\bar t)}{n_{0}} =
\exp\!\left(-\int_{c_{\mathrm{in}}(0)}^{\,c_{\mathrm{in}}(\bar{t})}
\bar{\Gamma}_{\mathrm p}(c)\,\frac{\mathrm{d}c}{c}\right)
\left[1-\frac{1}{\mathcal{K}_\text{linear}(\bar t)}
\left(1-\exp{\left[-\mathcal{K}_\text{linear}(\bar t)\right]}\right)\right],
\quad \bar t\in [0,\bar T],
\label{eq:n_ave_Linear_Approx_CSC}
\end{equation}
where \(\mathcal{K}_\text{linear}(\bar t)=\dfrac{\bar\Gamma_{\mathrm p}(c_{\mathrm{in}}(\bar t))}{\bar D_{\mathrm p}}\left(\bar t+\dfrac{\bar T}{\beta-1}\right)^{-1}\) with \(\beta\equiv c_1/c_0\).

An approximation for the final residual follows by evaluating \eqref{eq:n_ave_Linear_Approx_CSC} at \(\bar t=\bar T\):
\begin{equation}
\frac{N_\text{F}}{N_{0}} \approx \frac{n_{\text{ave}}(\bar T)}{n_{0}} =
\exp\!\left(-\int_{c_0}^{\,c_1} \bar{\Gamma}_{\mathrm p}(c)\,\frac{\mathrm{d}c}{c}\right)
\left[1-\frac{\bar D_{\text{p}}}{\bar \Gamma_{\text{p}}(c_1)}\,
\frac{\beta \bar T}{\beta -1}\left(1-\exp\!\left(-\frac{\bar \Gamma_{\text{p}}(c_1)}{\bar D_{\text{p}}}\,
\frac{\beta -1}{\beta \bar T}\right)\right)\right].
\label{eq:final_residual_CSC}
\end{equation}

We compare these predictions with simulations in figure~\ref{fig:CSC}. In panel (a), the approximation \eqref{eq:final_residual_CSC} (yellow dashed) is plotted against the numerical results (diamonds). In panel (b), the curves from \eqref{eq:n_ave_Linear_Approx_CSC} (solid lines) for \(T/\tau_\text{s}\in\{10, 100, 1000\}\) agree closely with the corresponding simulations (symbols).

Within the solute-concentration range we probed, variation in diffusiophoretic mobility did not alter our conclusions about the relative effects of diffuse versus sharp solute fronts. Nevertheless, a systematic analysis of regimes where the diffusiophoretic mobility approaches the particle diffusivity is needed to test the generality of these conclusions over a broader concentration range \citep{keh2000diffusiophoretic,prieve_diffusiophoresis_2019, gupta_diffusiophoresis_2020,ault_physicochemical_2024}.

\section[]{Electrolytes versus non-electrolytes}
\label{sec:ThreeCases}

While our model in \S\ref{sec:DEP_Gradual} applies for either sign of the diffusiophoretic mobility \(\Gamma_{\mathrm p}\), our measurements and simulations above focused on \(\beta>1\) with \(\Gamma_{\mathrm p}>0\), for which a slowly varying (“diffuse’’) solute concentration at the inlet (larger \(T\)) removes more particles and leaves a smaller residual particle fraction than an abrupt change in solute concentration. This raises a question on whether our observations are limited to positive mobilities and stem from the ``log-sensing" effect \citep{palacci_osmotic_2012,banerjee_soluto-inertial_2016,banerjee_long-range_2019,raj2023two}.

\begin{figure*}
    \centering
    \includegraphics[width=\textwidth, keepaspectratio]{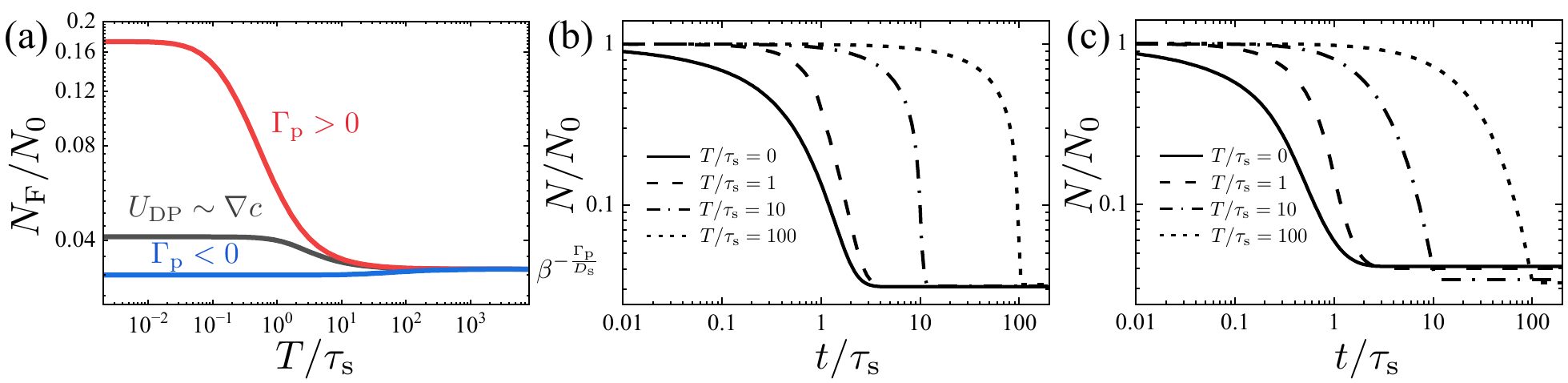}
    \captionsetup{width=\textwidth, justification=justified, singlelinecheck=false}
    \caption{\label{fig:ThreeCases}
    Comparison of electrolyte (\(\Gamma_\mathrm{p}>0\) and \(\Gamma_\mathrm{p}<0\)) and non-electrolyte (\(\Gamma_\mathrm{NE}\)) cases.
    (a) Final residual fraction \(N_\mathrm{F}/N_0\) versus the normalized solute transition time \(T/\tau_\mathrm{s}\), obtained from particle tracking with non-diffusive particles. Red: electrolyte with \(\Gamma_\mathrm{p}>0\); blue: electrolyte with \(\Gamma_\mathrm{p}<0\); black: non-electrolyte model \(U_\mathrm{DP}=\Gamma_\mathrm{NE}\nabla(c/c_0)\).
    (b–c) Temporal evolution of the residual fraction \(N/N_0\) for the \(\Gamma_\mathrm{p}<0\) electrolyte (b) and the non-electrolyte (c) cases at \(T/\tau_\mathrm{s}\in\{0,1,10,100\}\), computed from non-diffusive particle trajectories.
    Across cases, larger \(T/\tau_\mathrm{s}\) corresponds to more diffuse inlet ramps.
    }
\end{figure*}

To address this question, here, we first examine the case \(\beta<1\) with constant \(\Gamma_{\mathrm p}<0\) for non-Brownian particles (negligible \(D_{\mathrm p}\)) driven by linearly varying solute concentration at the inlet.
The two data sets, shown together in figure~\ref{fig:ThreeCases}(a), reveal three trends:
\begin{enumerate}
\item at fixed \(|\ln\beta|\) and \(|\Gamma_{\mathrm p}|\), extraction is stronger for \(\Gamma_{\mathrm p}<0\) than for \(\Gamma_{\mathrm p}>0\), 
\item in the \(\Gamma_{\mathrm p}<0\) regime, increasing the ramp duration \(T\) (i.e. a more diffuse front) \emph{slightly} weakens the extraction, and \item for both signs of \(\Gamma_{\mathrm p}\), when \(T/\tau_{\mathrm s}\gg1\) the residual particle fraction approaches the asymptotic prediction \(\beta^{-\Gamma_{\mathrm p}/D_{\mathrm s}}\) from equation~\eqref{eq:General_N_Ave_ND} in \S\ref{sec:DEP_Gradual}.
\end{enumerate}
These trends are consistent with the observations of \citet{migacz_enhanced_2024}.
The temporal evolution for the \(\Gamma_{\mathrm p}<0\) cases is shown in Fig.~\ref{fig:ThreeCases}(b).
Compared with the \(\Gamma_{\mathrm p}>0\) case (Fig.~\ref{fig:NF_T}(a)), the residual \(N/N_0\) initially decreases more slowly but then drops rapidly near the end of the ramp.
This acceleration follows from the functional form of the diffusiophoretic velocity \(U_{\mathrm{DP}}=\Gamma_{\mathrm p}\nabla\ln c=\Gamma_{\mathrm p}\,\nabla c / c\):
for \(\Gamma_{\mathrm p}<0\) we have \(\beta<1\) and the ramp drives \(c\) downward, so the factor \(1/c\) grows in time, amplifying \(|\nabla c|/c\) and hence the late-time drift.
Finally, for non-electrolytes, where the diffusiophoretic velocity takes the form \(U_\text{DP}=\Gamma_\text{NE}\nabla (c/c_0)\), we find that a diffuse solute front consistently removes more particles than a sharp step.\\

\noindent \textbf{Sign asymmetry from the solute equation written in \(\ln c\):}
To rationalise these trends, recall that the diffusiophoretic velocity
\(U_{\mathrm{DP}}=\Gamma_{\mathrm p}\,\boldsymbol{\nabla}\ln (c/c_0)\)
is controlled by the gradient of the effective concentration
\(e\equiv\ln(c/c_0)\).
Dividing the solute-transport equation~\eqref{eq:3} by \(c\) gives
\begin{equation}
  \partial_t e - D_{\mathrm s}\,\partial_{xx}e
  = D_{\mathrm s}\bigl(\partial_x e\bigr)^2,
  \label{eq:E_eqn}
\end{equation}
with \(e(x,0)=0\), the inlet ramp

\begin{equation}
e(0,t)=
\begin{cases}
\ln\!\bigl[1+(\beta-1)t/T\bigr], & 0\le t<T,\\
\ln\beta, & t\ge T,
\end{cases}
\end{equation}

and \(\partial_x e(L,t)=0\).

Because \(|U_{\mathrm{DP}}|\propto |\partial_x e|\), the maximum solute concentration contrast along the pore \(|\Delta e_{\max}|=|\ln\beta|\) serves as the fixed available “driving budget’’ for extraction. However, \eqref{eq:E_eqn} is \emph{non-conservative}: the positive source term \(D_{\mathrm s}(\partial_x e)^2\) generates effective concentration in the interior for \emph{any} \(e(x,t)\) field.
Its influence, however, depends on the sign of the target level \(\ln\beta\):
\begin{itemize}
\item For \(\Gamma_{\mathrm p}>0\) and \(\beta>1\), effective concentration \(e\) increases from \(0\) to \(+\ln\beta\):
the source \emph{assists} equilibration toward the inlet value, shortening the lifetime of gradients.
\item For \(\Gamma_{\mathrm p}<0\) and \(\beta<1\), effective concentration \(e\) decreases from \(0\) to \(-\ln\beta\):
the same positive source now \emph{opposes} the downward relaxation, keeping gradients alive for longer.
\end{itemize}

In short, the positive source in~\eqref{eq:E_eqn} accelerates convergence when the target \(e\) is positive and impedes it when the target is negative.
This sign-asymmetric interplay could explain the stronger overall extraction for \(\Gamma_{\mathrm p}<0\) at fixed \(|\ln\beta|\).\\

\noindent \textbf{Non–electrolyte case \(\big(U_\text{DP}=\Gamma_\text{NE}\,\nabla(c/c_0)\big)\):}
For non–electrolytes the diffusiophoretic velocity is
\(U_\text{DP}=\Gamma_\text{NE}\,\nabla(c/c_0)\), with \(c_0\) a reference concentration \citep{anderson_motion_1982,anderson_colloid_1989,velegol_origins_2016,williams_diffusioosmotic_2020,shim_motorless_2024}.
Unlike the electrolyte case written in terms of \(e=\ln(c/c_0)\), the solute equation expressed in \(c\) has no quadratic source term analogous to \(D_\mathrm{s}(\partial_x e)^2\); therefore the sign–asymmetric argument does not apply directly.

For slowly varying inlet histories, an exact counterpart of the analysis in \S\ref{sec:DEP_Gradual} can be developed. Proceeding as there, the particle density obeys
\begin{equation}
\frac{\partial n}{\partial \bar{t}}
+\frac{\partial}{\partial \bar{x}}
\!\left[\,
\bar{\Gamma}_\mathrm{NE}\!\big(c_\text{in}(\bar t)\big)\,\frac{(\bar x-1)\,c_\text{in}'(\bar t)}{c_0}\,n
\right]
=\epsilon\,\frac{\partial^2 n}{\partial \bar{x}^2},
\label{eq:nfield_approx_eqn_NE}
\end{equation}
where \(\bar{\Gamma}_\mathrm{NE}\equiv \Gamma_\text{NE}/D_\mathrm{s}\) and \(\epsilon=\bar D_\mathrm{p}\).
Using the same singular–perturbation (inner/outer) construction yields
\begin{equation}
\frac{n(\bar x,\bar t)}{n_0}
=\exp\!\left(
-\!\int_{c_{\mathrm{in}}(0)}^{\,c_{\mathrm{in}}(\bar t)}
\bar{\Gamma}_{\mathrm{NE}}(c)\,\frac{\mathrm{d}c}{c_0}
\right)
\left[1-\exp\!\big(-\mathcal{K}_{\mathrm{NE}}(\bar t)\,\bar x\big)\right],
\label{eq:nField_composite_NE}
\end{equation}
with diffusion–layer parameter
\(\mathcal{K}_{\mathrm{NE}}(\bar t)=\dfrac{\bar{\Gamma}_{\mathrm{NE}}\!\big(c_\text{in}(\bar t)\big)}{\bar D_{\mathrm p}}\,
\dfrac{c_\mathrm{in}'(\bar t)}{c_0}\).
Averaging over the pore gives
\begin{equation}
\frac{n_{\text{ave;NE}}(\bar t)}{n_0}
=\exp\!\left(
-\!\int_{c_{\mathrm{in}}(0)}^{\,c_{\mathrm{in}}(\bar t)}
\bar{\Gamma}_{\mathrm{NE}}(c)\,\frac{\mathrm{d}c}{c_0}
\right)
\left[
1-\frac{1}{\mathcal{K}_{\mathrm{NE}}(\bar t)}
\Big(1-e^{-\mathcal{K}_{\mathrm{NE}}(\bar t)}\Big)
\right].
\label{eq:General_N_Ave_NE}
\end{equation}

In the non–diffusive limit \(\bar D_\mathrm{p}\to 0\) and for constant mobility \(\bar{\Gamma}_{\mathrm{NE}}\), the final residual fraction is
\begin{equation}
\frac{N_{\text{F;NE}}}{N_0}
=\lim_{\bar D_\mathrm{p}\to 0}\frac{n_{\text{ave;NE}}(\infty)}{n_0}
=\exp\!\left[-\,\bar{\Gamma}_{\mathrm{NE}}\,
\frac{c_\text{in}(\infty)-c_\text{in}(0)}{c_0}\right].
\label{eq:General_N_Ave_ND_NE}
\end{equation}

We also simulated the non-electrolyte case with constant \(\Gamma_{\mathrm{NE}}\), using non-Brownian particles subjected to linear inlet ramps of varying duration \(T\).
For ease of comparison, we set \(c_0=c_{\mathrm{in}}(0)\) and choose
\(\Gamma_{\mathrm{NE}}=\Gamma_{\mathrm p}\,\dfrac{\ln\beta}{\beta-1}\) so that, in the limit \(T/\tau_{\mathrm s}\to\infty\), the asymptotic residual matches the electrolyte result:
\(N_\mathrm{F}/N_0\to \beta^{-\Gamma_{\mathrm p}/D_{\mathrm s}}
=\exp[-\bar\Gamma_{\mathrm{NE}}(\beta-1)]\) with \(\bar\Gamma_{\mathrm{NE}}=\Gamma_{\mathrm{NE}}/D_{\mathrm s}\).
The outcomes are shown in Fig.~\ref{fig:ThreeCases}(a). As \(T/\tau_{\mathrm s}\) increases, \(N_\mathrm{F}/N_0\) decreases and approaches the same asymptote, mirroring the trend for \(\Gamma_{\mathrm p}>0\).
The corresponding time evolutions likewise resemble the \(\Gamma_{\mathrm p}>0\) case (compare figure.~\ref{fig:ThreeCases}(c) with figure.~\ref{fig:NF_T}(a)).

\section{\label{sec:Discussion}Discussion}

Our work probes a fundamental question on the role of diffusiophoresis in the transport of colloids in porous media: does dispersion of the solute front weaken the phoretic effects? Combining microfluidic experiments and numerical simulations, we showed that while the solute front becomes more dispersed as it travels through the medium, the residual particle fraction plateaus to the same value in different fields of view along the medium. This observation suggests that perhaps surprisingly the dispersion of solute front does not weaken the overall diffusiophoretic migration from low permeability pockets of the medium.

To gain further insight, we then probed the phoretic migration of colloids from 1D dead-end pores, varying the dispersion of the solute front. Our microfluidic experiments showed that dispersed fronts perhaps paradoxically enhance the removal efficiency of the particles compared to sharp fronts. Using numerical simulations and analytical modeling, we then showed that this observation is due to the persistence of solute gradients in the dispersed front case, leading to a more uniform removal of particles. Our analytical model shows that the final residual particle fraction is mainly a function of solute evolution time $\bar{T} = T/\tau_{\text{s}}$. 

Applying this insight to the porous medium studied in Sec.~\ref{sec:PM}, where the area-weighted average pore length is approximately $L \approx 130~\mu\text{m}$ and $\tau_{\text{s}} \approx 12~\text{s}$, we find an equivalent $\bar{T}$ of about $4$ to $6$. The dead-pore analysis predicts a final residual particle fraction of $\approx 0.04$ for this $\bar{T}$. Using the measured fraction of the dead end pore areas in the porous medium ($\eta = 0.03$) and the mean particle density after flushing the porous medium  ($\theta_{\text{F}}^{\text{A}} \approx 0.001$ in Fig.~1(d)), we can calculate the average final residual particle fraction $\alpha = \theta_{\text{F}}^{\text{A}}/\eta \approx 0.03$, which agrees well with our estimated value of $\alpha \approx 0.04$ for a single dead-end pore. \\

Our work therefore shows that, contrary to the intuitive expectation that sharper solute fronts should be more effective in driving the diffusiophoretic migration of colloids, smoothing the solute concentration gradient improves the extraction of particles from dead end pores. Our work generalizes the analysis of \citet{migacz_enhanced_2024}, who considered non-Brownian particles under linear inlet ramps and obtained closed-form expressions for trajectories and residual fractions. We extend that framework in two directions. First, we incorporate finite particle diffusivity, capturing diffusion-driven corrections that become relevant at large \(T/\tau_{\text{s}}\). Second,  we present a general formulation that accommodates a broad class of slowly varying inlet solute concentration profiles, including error-function–type transitions (Appendix~A.2). In the limiting case of negligible \(D_{\mathrm p}\), our predictions reduce to those of \citet{migacz_enhanced_2024}, ensuring consistency between the two approaches.\\

In the current descriptions of colloid transport in porous media, typically hydrodynamic and non-hydrodynamic DLVO-type interactions between colloids and surrounding matrix are considered \citep{ryan_colloid_1996,kretzschmar_mobile_1999,bradford_physical_2002,torkzaban_resolving_2007,bradford_colloid_2008,molnar_predicting_2015,morales_stochastic_2017,molnar_colloid_2019,miele_stochastic_2019,bizmark_multiscale_2020,perez_morphology_2020,patino_relating_2023,fan_anomalous_2022,bordoloi_structure_2022,al-zghoul_effects_2024,volponi_interception_2025}. While the solute concentration is known to influence the range of equilibrium DLVO-type interactions \citep{liu_colloid_1995,roy_chemical_1997,elimelech_relative_2000}, the phoretic effects driven by solute gradients have been mostly ignored so far. Our work demonstrates that, while these phoretic velocities can be much weaker than the typical background flow velocities, they remain persistent due to solute dispersion. Our work therefore builds on the growing evidence that solute gradients can have a strong effect on the diffusiophoretic transport of colloids and emulsions in porous media \citep{doan_confinement-dependent_2021,sambamoorthy_diffusiophoresis_2023,jotkar_impact_2024,jotkar_diffusiophoresis_2024,sambamoorthy_diffusiophoresis_2025,alipour_diffusiophoretic_2024}. Our findings could therefore have implications for controlling particle transport in a broad range of applications, from microfluidics and drug delivery to contaminant remediation and oil recovery in subsurface flows, where tuning the solute dispersion could be used to optimize particle extraction efficiency.

\backsection[Declaration of interests]{The authors report no conflict of interest.}

\appendix
\setcounter{figure}{0}
\renewcommand{\thefigure}{A\arabic{figure}}

\section{Validation of analytical model}\label{appA}

\subsection{Field expression validation}

\begin{figure*}
     \centering
     \includegraphics[width=\textwidth, keepaspectratio]{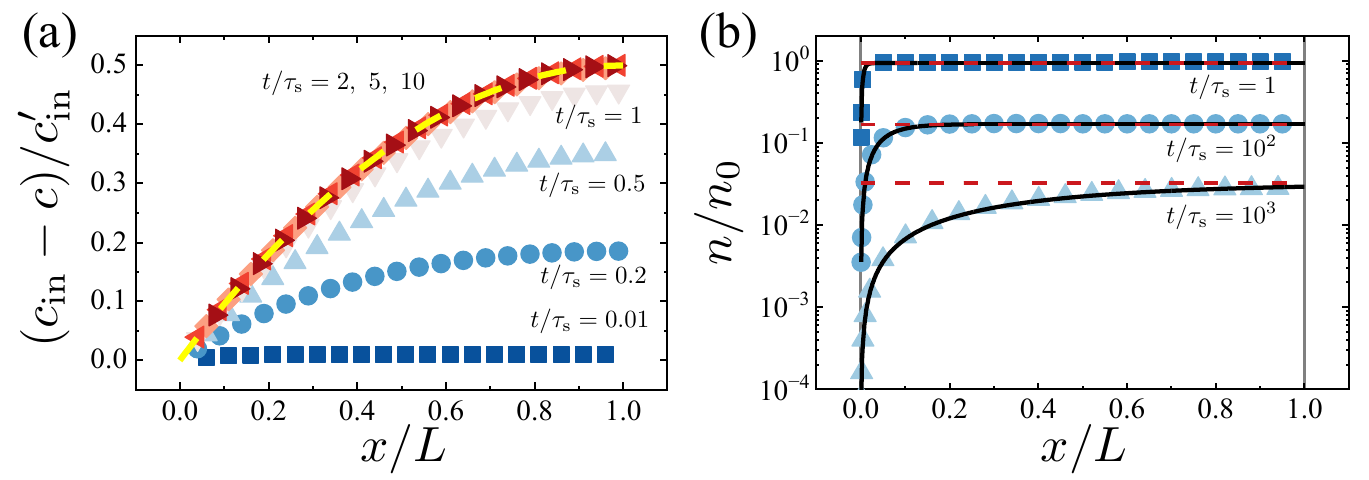}
     \captionsetup{width=\textwidth, justification=justified, singlelinecheck=false} 
     \caption{\label{fig:Gradual_Linear_Fields} 
     Analytical and numerical results for the evolution of solute and particle fields in a dead-end pore with large normalized solute transition times, $T/\tau_\text{s}$. (a) Evolution of the solute concentration profile for $T/\tau_\text{s} = 10$. The normalized solute concentration $(c - c_\text{in}) / c_\text{in}'$ (symbols) converges to the analytical solution $\frac{1}{2}(2\bar x - \bar x^2)$ (yellow dashed line) for $t/\tau_\text{s} > 1$. (b) Evolution of the particle density profile for $T/\tau_\text{s} = 10^3$. Symbols show numerical solutions of the advection--diffusion equation, while solid black lines denote the analytical predictions (Eq.~\eqref{eq:nField_Linear_Approx}). The red dashed lines represent the non-diffusive particle model, which diverges from numerical results at later times due to diffusion effects.}
\end{figure*}

We evaluated the accuracy of our approximations for both the solute and particle fields. It is found that the approximate solute field becomes valid with a relatively small $T/\tau_{\text{s}}$. Figure~\ref{fig:Gradual_Linear_Fields}(a) displays the normalized solute field $(c - c_\text{in})/c_\text{in}'$ for $T/\tau_{\text{s}} = 10$ at various times. As predicted by Eq.~(\ref{eq:c_field_approximate}), the field converges to the analytical form $\tfrac{1}{2}(2\bar{x} - \bar{x}^2)$ (yellow dashed line) once $t/\tau_{\text{s}} > 1$.

In contrast, the theoretically predicted particle density expression requires a much greater $T/\tau_{\text{s}}$ to be accurate. Figure~\ref{fig:Gradual_Linear_Fields}(b) shows the particle density field at different times for $T/\tau_{\text{s}} = 10^3$. Symbols represent numerical solutions to the advection--diffusion equation, while the solid black lines correspond to the predictions of Eq.~(\ref{eq:nField_Linear_Approx}). The good agreement confirms the validity of our analytical model under conditions of large solute transition times. During the particle extraction process, the particle density field remains nearly uniform, a hallmark of scenarios with slowly varying inlet solute profiles, except near the inlet, where a pronounced drop is induced by particle diffusion. For comparison, the red dashed lines show the non-diffusive particle model predictions, which deviate from the numerical results at later times due to the omission of diffusion effects.

\subsection{Error function type slowly varying solute profile with constant \(\bar{\Gamma}_\text{p}\)}

Having verified our framework for linear solute profiles, we now consider more realistic inlet conditions. In many porous media, dispersion often yields inlet solute concentration profiles that closely resemble an error function. Such profiles arise as solutions of the one-dimensional advection–diffusion equation. Here, we examine particle extraction under an error function type inlet solute profile:
\begin{equation}
c_\text{in}(\bar t) = c_{0} + \tfrac{1}{2}(c_{1} - c_{0})\left(1 + \text{erf}\left(\frac{2\bar t}{\bar \tau_\text{erf}} - \delta\right)\right),
\label{eq:error_function_solute_profile}
\end{equation}
where $\bar \tau_\text{erf} = \tau_\text{erf}/\tau_{\text{s}}$ sets the timescale of solute concentration change, and $\delta=3$ is a shift chosen so that $c_\text{in}(0)$ remains close to $c_{0}$.

To ensure that the inlet solute concentration is slowly varying and the thin diffusion layer condition~(\ref{eq:condition2}) is satisfied, enabling the application of our analytical model, we require $\bar \tau_\text{erf} \gg 1$ and
\begin{equation}
\frac{\bar \Gamma_{\text{p}}}{\bar D_{\text{p}}} \gg \frac{c_\text{in}(\bar{t})}{c_\text{in}'(\bar{t})} 
= \frac{\bar \tau_\text{erf}}{\beta - 1}\exp\left[\left(\frac{2\bar t}{\bar \tau_\text{erf}} - \delta\right)^2\right]\left(1 + \tfrac{1}{2}(\beta - 1)\left[1 + \text{erf}\left(\tfrac{2\bar t}{\bar \tau_\text{erf}} - \delta\right)\right]\right),
\label{eq:21}
\end{equation}
where $\beta = c_1/c_0$. Although appropriate choices of $\bar \tau_\text{erf}$ and $\beta$ can satisfy the second inequality, the first inequality may be violated at early and late times, when the inlet solute concentration is nearly constant. In these regimes, we approximate transport as pure diffusion (see Eq.~(\ref{eq:pure_diffusion})).

As a result, the extraction process is divided into three stages: pure diffusion (Eq.~(\ref{eq:pure_diffusion})) at early times, combined diffusiophoresis and diffusion (Eq.~(\ref{eq:General_N_Ave})) at intermediate times, and pure diffusion (Eq.~(\ref{eq:pure_diffusion})) again once the solute gradient has effectively vanished. We determine the two transition points, $\bar t_\text{c1}$ and $\bar t_\text{c2}$, between these stages using the criterion
\begin{equation}
\frac{c_\text{in}(\bar t_{\text{c}i})}{c_\text{in}'(\bar t_{\text{c}i})} = \frac{r_i}{\bar \tau_\text{erf}}
\cdot \frac{\bar \Gamma_{\text{p}}}{\bar D_{\text{p}}},
\end{equation}
where $r_i \ll 1$ ($i=1,2$) are fitting parameters. Here, we choose $r_1 = 6\times 10^{-4}$ and $r_2 = 8\times 10^{-4}$.

Under these conditions, the average particle density evolves as
\begin{equation}
\frac{n_\text{ave}(\bar t)}{n_0} =
\begin{cases}
\displaystyle \sum_{m=0}^{\infty}\frac{2}{\lambda_m^2} \exp(-\lambda_m^2 \bar D_\text{p} \bar t), & \bar t \in [0,\bar t_\text{c1}], \\[10pt]

\displaystyle \frac{n_\text{ave}(\bar t_\text{c1})}{n_0}\left(\frac{c_{\text{in}}(\bar t)}{c_{\text{in}}(\bar t_\text{c1})}\right)^{-\bar \Gamma_\text{p}} 
\frac{\left[1 - \left(1 - 1/K(\bar t)\exp\bigl(-K(\bar t)\bigr)\right)\right]}{\left[1 - \left(1 - 1/K(\bar t_\text{c1})\exp\bigl(-K(\bar t_\text{c1})\bigr)\right)\right]}, & \bar t \in (\bar t_\text{c1}, \bar t_\text{c2}], \\[10pt]

\displaystyle \frac{n_\text{ave}(\bar t_\text{c2})}{n_0}\sum_{m=0}^{\infty}\frac{2}{\lambda_m^2}\exp[-\lambda_m^2 \bar D_\text{p} (\bar t - \bar t_\text{c2})], & \bar t \in (\bar t_\text{c2}, \infty),
\end{cases}
\label{eq:Composite_Solution_K}
\end{equation}
where $\lambda_m = \tfrac{(2m+1)\pi}{2}$ and $K(\bar t) = \frac{\bar \Gamma_{\text{p}}}{\bar D_{\text{p}}}\frac{c_\text{in}'(\bar{t})}{c_\text{in}(\bar{t})}$.

\begin{figure*}
    \centering
    \includegraphics[width=\textwidth, keepaspectratio]{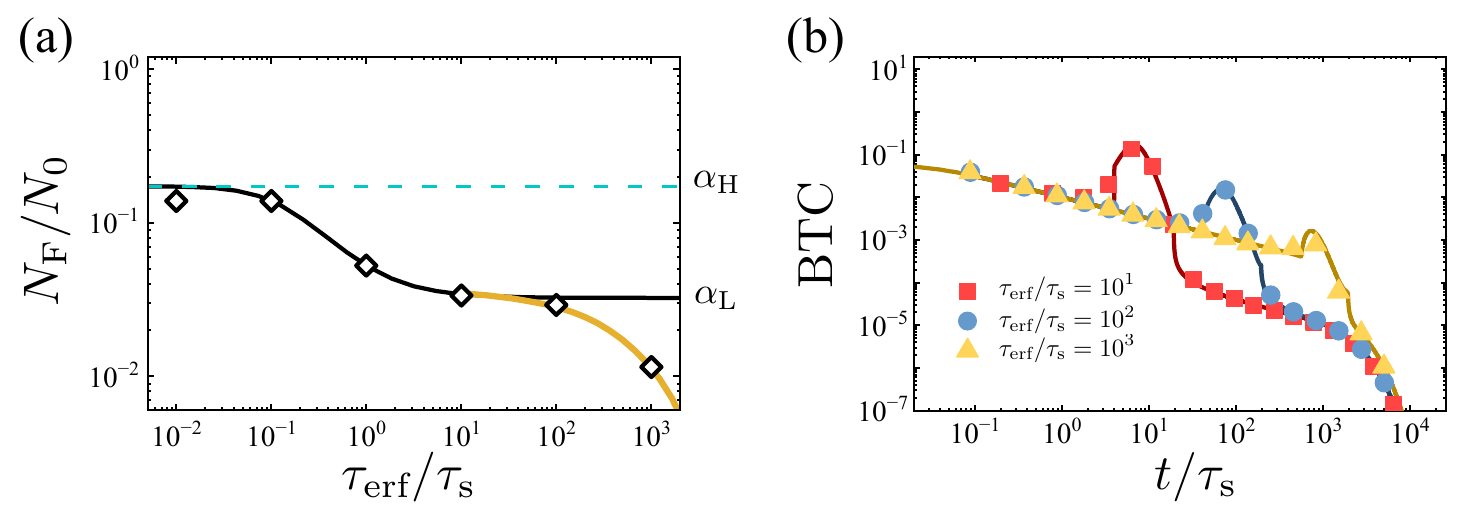}
    \captionsetup{width=\textwidth, justification=justified, singlelinecheck=false} 
    \caption{\label{fig:Gradual_Error_BTC} 
    Validation of our analytical framework for an error function type slowly varying solute profile.
    (a) Final residual particle fraction, defined as the value when the average solute concentration in the dead-end pore reaches $99\%$, as a function of the effective solute transition time. Results are obtained from continuous solutions for diffusive particles (diamond symbols) and from particle tracking under the non-diffusive assumption (black solid lines). The yellow solid line represents our analytical prediction that accounts for both diffusiophoresis and particle diffusion.
    (b) Breakthrough curves for $\tau_\text{erf}/\tau_{\text{s}} = 10^1, 10^2,$ and $10^3$. Symbols represent numerical results, while solid lines show our analytical predictions, demonstrating good agreement over a wide range of solute transition times.
    }
\end{figure*}

The piecewise definition reflects the three distinct stages of the extraction process: early time pure diffusion, intermediate time coupled diffusiophoresis and diffusion, and late time pure diffusion.

Figure~\ref{fig:Gradual_Error_BTC}(a) shows the final residual particle fraction (evaluated when the average solute concentration in the dead-end pore reaches 99\% of its final value) as a function of $\tau_\text{erf}/\tau_\text{s}$. The black solid line corresponds to particle tracking results under the non-diffusive assumption, while the diamond symbols represent solutions of the advection–diffusion equation that include particle diffusion. Similar to the linear case, we identify a highest residual fraction $\alpha_{\text{H}}=0.173$ and a lowest residual fraction $\alpha_{\text{L}}=0.032$. As $\tau_\text{erf}/\tau_\text{s}$ increases, the diffusive model diverges from the non-diffusive one, and our analytical prediction successfully captures the further reduction in the residual fraction introduced by particle diffusion.

Figure~\ref{fig:Gradual_Error_BTC}(b) shows breakthrough curves (BTCs) for error function type solute profiles at various $\tau_\text{erf}/\tau_\text{s}$ values. Symbols represent numerical solutions of the advection–diffusion equations, and solid lines represent our theoretical predictions. The good agreement confirms the reliability of our three stage model for the error function type slowly varying solute profile.

\newpage

\subsection{Influence of salt type: LiCl versus NaCl}

We repeated the dead-end pore experiments described in section~\ref{sec:DEP_Linear} using NaCl as the solute. Because colloids in NaCl can adsorb onto channel walls at high salt concentrations, we used lower concentrations: $c_0=0.1~\textrm{mM}$ for the initial suspension and $c_1=10~\textrm{mM}$ for the introduced colloid-free solution. For direct comparison, we repeated the experiments with LiCl at the same concentrations ($c_0=0.1~\textrm{mM}$, $c_1=10~\textrm{mM}$).

\begin{figure*}
    \centering
    \includegraphics[width=\textwidth, keepaspectratio]{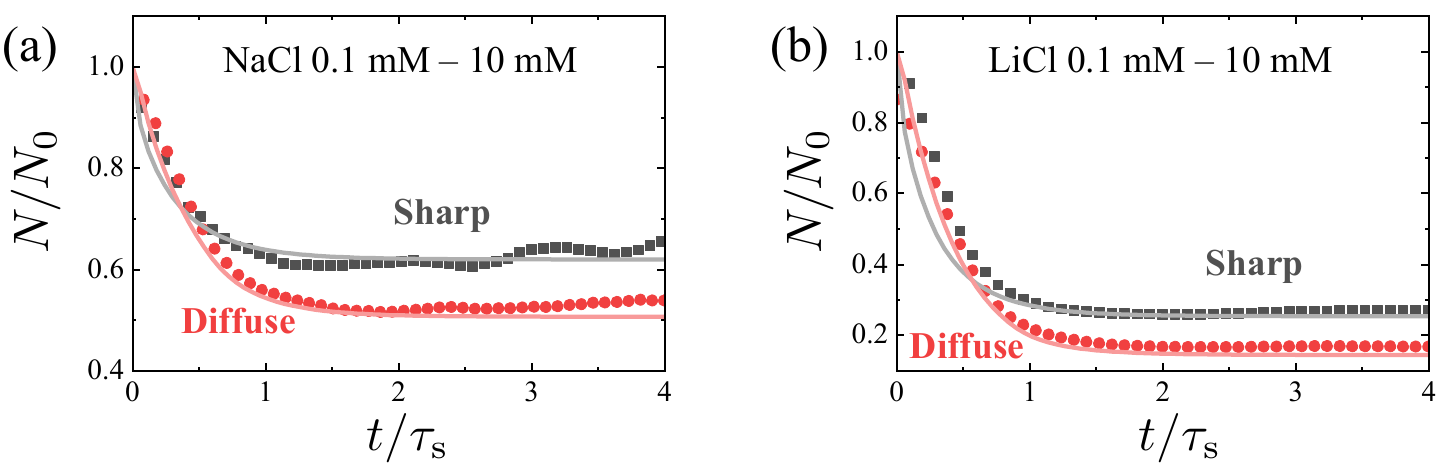}
    \captionsetup{width=\textwidth, justification=justified, singlelinecheck=false} 
    \caption{\label{fig:NaCl_DEP}
    Dead-end pore experiments with (a) NaCl and (b) LiCl with \(c_0=0.1~\mathrm{mM}\) and \(c_1=10~\mathrm{mM}\).
Symbols show experimental measurements; solid lines are fits from the non-diffusive particle-trajectory model in Eq.~\eqref{eq:Udp_Expression1} using a constant diffusiophoretic mobility.
The fitted mobilities are \(\Gamma_\mathrm{p}^{\mathrm{NaCl}}\approx0.4\times10^{-9}~\mathrm{m^2\,s^{-1}}\) and \(\Gamma_\mathrm{p}^{\mathrm{LiCl}}\approx0.7\times10^{-9}~\mathrm{m^2\,s^{-1}}\).
    }
\end{figure*}

We observe that for both electrolytes, a diffuse solute front extracts more particles from the dead-end pore than a sharp front, as demonstrated in Fig.~\ref{fig:NaCl_DEP}, where symbols denote experimental measurements, and solid lines are fits from the non-diffusive particle-trajectory model (Eq.~\ref{eq:Udp_Expression1}) with a constant diffusiophoretic mobility. The fitted mobilities are \(\Gamma_\mathrm{p}^{\mathrm{NaCl}}\approx 0.4\times10^{-9}~\mathrm{m^2\,s^{-1}}\) and \(\Gamma_\mathrm{p}^{\mathrm{LiCl}}\approx 0.7\times10^{-9}~\mathrm{m^2\,s^{-1}}\), with NaCl exhibiting the lower value.

\subsection{Comparison between CSC and constant–\(\Gamma_\text{p}\) models}
\label{subsec:CSC_vs_const}

We compare 2D dead–end–pore–with–channel simulations (see \S\ref{sec:DEP_Linear}) using two expressions for the diffusiophoretic mobility: (i) a constant \(\Gamma_\text{p}\), already described in \S\ref{sec:DEP_Linear}, and (ii) a concentration–dependent mobility given by the constant–surface–charge (CSC) model (\S\ref{sec:cEffect}). In the CSC runs, the slope \(a_1\) in the empirical \(\zeta(c)\) relation is treated as a single fitting parameter: it is calibrated to match the sharp–front experiment, yielding \(a_{1}\approx -42.0~\mathrm{mV\,decade^{-1}}\). The diffuse–front simulation then uses the same \(a_1\) and the inlet history inferred from the fluorescein signal with time–rescaled as in \S\ref{sec:DEP_Linear}.

\begin{figure*}
    \centering
    \includegraphics[width=\textwidth, keepaspectratio]{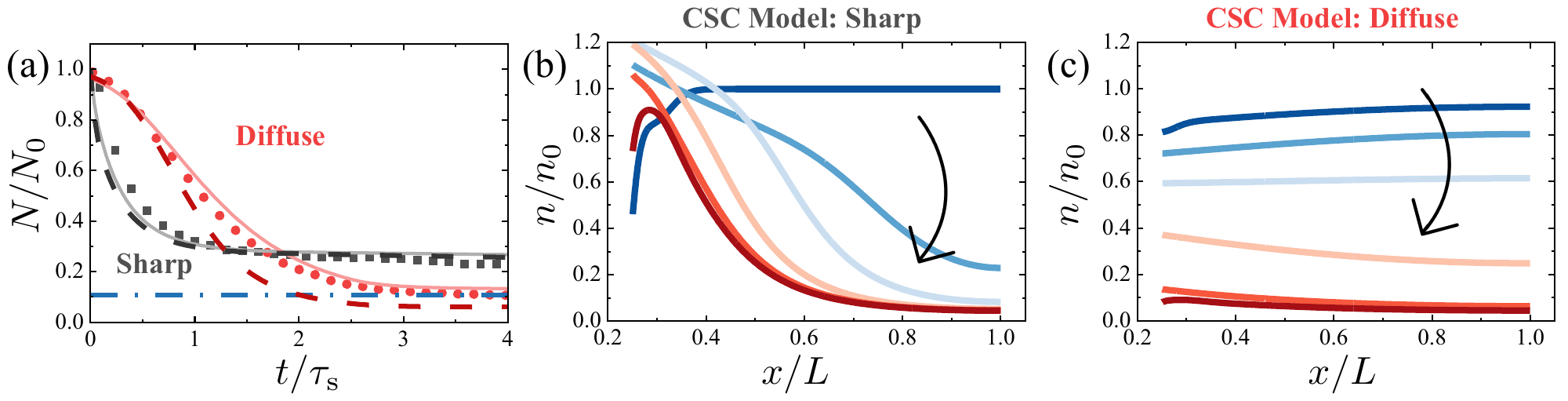}
    \captionsetup{width=\textwidth, justification=justified, singlelinecheck=false} 
    \caption{\label{fig:CSC_Comparison}
    Comparison of simulations using the CSC mobility model versus a constant \(\Gamma_\text{p}\).
    (a) Time evolution of the residual fraction \(N/N_0\).
    Symbols: experiments; solid lines: 2D simulations with constant \(\Gamma_\text{p}\); dashed lines: 2D simulations with the CSC model.
    (b) Longitudinal particle–density profiles along the dead-end pore for the \emph{sharp} front, evaluated at \(t/\tau_{\mathrm s}=0,\,0.1,\,0.2,\,0.5,\,1,\,2\).
    (c) Longitudinal particle–density profiles for the \emph{diffuse} front, evaluated at \(t/\tau_{\mathrm s}=0,\,0.25,\,0.5,\,1,\,2,\,4\).
    The CSC model captures the qualitative trends and matches the sharp–front case closely, while slightly underestimating the diffuse–front residuals relative to experiment.
    }
\end{figure*}

Figure~\ref{fig:CSC_Comparison}(a) overlays the CSC results on the experimental data and the constant \(\Gamma_\text{p}\) simulations from figure~\ref{fig:DEP_EXP}(e). The CSC model reproduces the qualitative trends in both sharp and diffuse cases. Quantitatively, it closely matches the sharp–front residuals, while for the diffuse front it slightly underestimates the residual fraction relative to the measurements. Panels~\ref{fig:CSC_Comparison}(b–c) show the corresponding longitudinal density profiles along the pore. Compared with the experimental profiles in figure~\ref{fig:DEP_EXP}(f–g) and the constant \(\Gamma_\text{p}\) simulations in figure~\ref{fig:DEP_EXP}(i–j), the CSC model exhibits stronger inlet focusing in the sharp case, whereas for the diffuse case it likewise yields a nearly uniform profile at late times. Overall, for our dead-end pore experiments with LiCl concentration varying from \(c_0=1~\textrm{mM}\) to \(c_1=100~\textrm{mM}\), both constant and variable mobility models capture the key qualitative behaviors observed in the experiments. It would certainly be interesting to explore further the impact of variable zeta potential and phoretic mobility over a wider range of salt concentrations and salt types as recent works offer evidence for their potential role \citep{akdeniz_diffusiophoresis_2023, lee_role_2023}.

\newpage

\bibliographystyle{jfm}
\bibliography{jfm}

\end{document}